\documentstyle[epsfig]{ar}

\begin{document}



\newcommand{\nub}{\overline{\nu}}
\newcommand{\dmsq}{\delta M^2}
\newcommand{\optbar}[1]{\shortstack{{\tiny (\rule[.4ex]{1em}{.1mm})} 
  \\ [-.7ex] $#1$}}
\def\lsim{\raisebox{-.6ex}{$\stackrel{<}{\sim}$}}
\def\gsim{\raisebox{-.6ex}{$\stackrel{>}{\sim}$}}
\newcommand{\ra}{\mbox{$\rightarrow$}}
\newcommand{\Eq}[1]{Eq.~(\ref{eq#1})}
\newcommand{\beq}{\begin{equation}}
\newcommand{\eeq}{\end{equation}}
\newcommand{\numu}{\nu_{\mu}}
\newcommand{\nutau}{\nu_{\tau}}
\newcommand{\numubar}{\overline{\nu}_{\mu}}
\newcommand{\nuebar}{\overline{\nu}_{e}}
\newcommand{\nue}{\nu_{e}}
\newcommand{\Rmue}{R_{\mu/e}}
\newcommand{\Rdatamc}{R_{\mathrm{Data/MC}}}

\title{\begin{flushright} {\normalsize Preprint LNS-99-288, NSF-PT-99-1,
      UR-1569}\vspace{2ex}  \end{flushright}
Neutrino Mass and Oscillation}  
\markboth{P.~Fisher, B.~Kayser, K.S.~McFarland}{Neutrino Mass and Oscillation}

\author{Peter Fisher\affiliation{Massachusetts Institute of Technology,
    Department of Physics, Cambridge, MA 02192}
Boris Kayser\affiliation{National Science Foundation, Division of Physics,
Arlington, VA 22230}
Kevin S.~McFarland\affiliation{University of Rochester, Department of Physics
                     and Astronomy, Rochester, NY 14627}}


\begin{abstract}
The question of neutrino mass is one of the major riddles in particle physics. 
Recently, strong evidence that neutrinos have nonzero masses has been found. While
tiny, these masses could be large enough to contribute significantly to the mass
density of the universe. The evidence for nonvanishing neutrino masses is based on
the apparent observation of neutrino oscillation---the transformation of a neutrino
of one type or ``flavor'' into one of another. 
We explain the physics of neutrino oscillation, and review and weigh the evidence
that it actually occurs in nature. We also discuss the constraints on neutrino mass
from cosmology and from experiments with negative results. After presenting
illustrative neutrino mass spectra suggested by the present data, we consider how
near- and far-future experiments can further illuminate the nature of neutrinos and
their masses.
\end{abstract}

\maketitle


\section{INTRODUCTION}  \label{sec1}

Neutrinos are among the most abundant particles in the universe. Indeed, even
within a single human being, there are more than $10^7$ neutrinos left over
from the Big Bang. In addition, passing through each person on earth every
second are some $10^{14}$ neutrinos from the sun, another $10^3$ neutrinos
made in the earth's atmosphere by cosmic rays, and additional neutrinos from
other natural and man-made sources. Evidently, neutrinos are very common in
the universe, and very commonly encountered by people.  Thus, it would be
very nice to know something about them.

It has long been known that neutrinos are very light. However, it has not
been known whether they are completely massless, or have small but nonzero
masses. The answer to this question has major consequences, both for physics
and for astrophysics. If the neutrino masses are very small but nonvanishing,
then they are probably caused by new physics beyond the realm of the highly
successful Standard Model of the weak and electromagnetic interactions.
Measured values of the masses would be a clue to the nature of the new
physics. In addition, because neutrinos are so abundant in the universe, even
tiny nonzero neutrino masses can result in an important neutrino contribution
to the mass density of the universe.

Recently, strong evidence that neutrinos do have nonzero masses has at
long last been found. This evidence stems from the apparent observation that
a neutrino of one type or ``flavor'', such as a muon neutrino, can metamorphose
into a neutrino of another flavor, such as a tau neutrino. This
metamorphosis, known as neutrino oscillation, implies neutrino mass, as we
shall discuss.

Three hints that neutrinos oscillate have been observed. The first---the
behavior of neutrinos produced in the earth's atmosphere by cosmic
rays---provides rather convincing evidence of oscillation. The second---the
observed fluxes of neutrinos from the sun---provides a strong further hint of
it. The third---the behavior of neutrinos in a beam at Los Alamos---provides
an additional but unconfirmed hint.

In this article, we will briefly review the physics of neutrino mass and
oscillation, and then discuss and critique the three hints that neutrinos
actually oscillate. Next, we will indicate the bounds placed on neutrino mass
by astrophysics and by various laboratory experiments with negative results.
Then, we will describe several illustrative neutrino mass scenarios suggested
by the observed oscillation hints and consistent with the existing bounds.
Finally, we will conclude by discussing the future steps which need to be
taken to test the three existing hints of oscillation, and to learn more
about the nature of neutrinos and their masses.

\section{THE EXPECTATION THAT NEUTRINOS HAVE MASS}   \label{sec2}

Over the years, numerous experiments have sought evidence of neutrino mass.
This extensive search has been driven, in part, by the fact that from a
theoretical perspective, it is natural to expect that neutrinos do have
nonvanishing masses. Unlike photons, neutrinos are not required to be
massless by gauge invariance or any other symmetry principle. Moreover,
neutrinos appear to be closely related to the charged leptons and the quarks,
all of which certainly have nonvanishing masses. In the Grand Unified
Theories, which unify the weak, electromagnetic, and strong interactions,
this relation takes the form of the placement of each neutrino in a multiplet
together with at least one charged lepton, one positively-charged quark, and
one negatively-charged quark. Apart from the neutrino, every member of this
multiplet is known to be massive. Thus the neutrino would have to be an
exceptional member of the multiplet to be massless.

Assuming that neutrinos do have mass, we have to understand why they are
nevertheless so much lighter than the charged leptons and quarks. The most
popular explanation of this fact is the ``see-saw mechanism'' \cite{b1}. To
understand how this mechanism works, let us recall that, unlike charged
particles, neutrinos may be their own antiparticles. A neutrino which is its
own antiparticle consists of just two states with a common mass: one with
spin up and one with spin down. Such a neutrino is called a Majorana
neutrino. By contrast, a neutrino which is distinct from its antiparticle
consists of four states with a common mass: the spin-up and spin-down
neutrino, plus the spin-up and spin-down antineutrino. This collection of
four states is called a Dirac neutrino.  In the see-saw mechanism, a
four-state Dirac neutrino ${\cal N}^D$ of mass $M^D$ gets split by ``Majorana
mass terms'' into a pair of two-state Majorana neutrinos. One of the latter
neutrinos, $\nu^M$, has a small mass $M_\nu$ and is identified as one of the
observed light neutrinos. The other, $N^M$, has a large mass $M_N$ reflecting
the high mass scale of some new physics beyond the Standard Model, and has
not been observed. The character of the breakup of ${\cal N}^D$ into $\nu^M$
and $N^M$ is such that $M_\nu M_N \cong M_D^2$.  Now, it is reasonable to
expect that the mass $M_D$ of the Dirac particle ${\cal N}^D$ is of the same
order as the typical mass, $M_{\ell \mbox{ \scriptsize or }q}$, of the
charged leptons $\ell$ and quarks $q$, since the latter are Dirac particles
as well. Then, $M_\nu M_N \sim M_{\ell \mbox{ \scriptsize or }q}^2$. With
$M_{\ell \mbox{ \scriptsize or }q}$ a typical charged lepton or quark mass
and $M_N$ very large, this ``see-saw relation'' explains why $M_\nu$ is very
small. Very importantly, the see-saw mechanism predicts that neutrinos are
Majorana particles.

In a typical model, the heavy Majorana neutral lepton $N^M$ participates in
some hypothetical feeble interaction beyond the familiar weak interaction.
However, it does not participate in the weak interaction itself. For this
reason, it (and any other neutrino which is free of normal weak interactions)
is sometimes referred to as a ``sterile neutrino''.


\section{NEUTRINO OSCILLATION}\label{sect:osc}

In the quest for evidence of neutrino mass, the experiments which currently
can reveal the smallest neutrino masses are those which search for neutrino
oscillation. Let us briefly recall the physics of this process, starting with
oscillation in empty space.

\subsection{Oscillation in Vacuum}   \label{ssec30}

We shall denote by ``$\ell$'' the charged lepton of flavor $\ell = e$, $\mu$,
or $\tau$. Associated with a charged lepton $\ell$ of a given flavor is a
neutrino, $\nu_\ell$, of the same flavor. The meaning of this association is
that when a neutrino of a given flavor interacts with matter and produces a
charged lepton, this charged lepton always has the same flavor as the
neutrino: $\nu_e$ produces an $e$, $\nu_\mu$ a $\mu$, and $\nu_\tau$ a
$\tau$. Similarly, when a reaction creates a charged lepton and an
accompanying neutrino, the latter two particles are of the same flavor.

If neutrinos have masses, then a neutrino of definite flavor, $\nu_\ell$,
need not be a mass eigenstate. Indeed, if leptons behave as quarks do, then
$\nu_\ell$ is a coherent superposition of mass eigenstates, given by 
\beq 
|\,\nu_\ell \rangle = \sum_m U_{\ell m} \, |\, \nu_m \rangle .
\label{eq1}
\eeq 
Here, the $\nu_m$ are the mass eigenstates. There are at least three of
them, and perhaps more. The coefficients $U_{\ell m}$ form a matrix $U$ known
as the leptonic mixing matrix. According to the Standard Model (SM), extended
to include neutrino masses, $U$ is unitary.

The fact that a neutrino of definite flavor is a superposition of several
mass eigenstates, whose differing masses cause them to propagate differently, 
leads to {\em neutrino oscillation
in vacuum}---the transformation of a neutrino of one flavor into one of a
different flavor as the neutrino moves through empty space.  The amplitude
for the transformation $\nu_\ell \ra \nu_{\ell^\prime}$ is given by 
\beq
A(\nu_\ell \ra \nu_{\ell^\prime}) = \sum_m A(\nu_\ell \mbox{ is }\nu_m)\,
A(\nu_m \mbox{ propagates}) \, A(\nu_m \mbox{ is }\nu_{\ell^\prime}) \;.
\label{eq2}
\eeq 
Here, $A$ denotes an amplitude. From \Eq{1}, $A(\nu_\ell \mbox{ is }\nu_m) =
U_{\ell m}$, and from its inverse, $A(\nu_m \mbox{ is }\nu_{\ell^\prime}) =
U_{\ell^\prime m}^*$. It is not difficult to show that if the neutrino has
energy $E$, if it propagates for a distance $L$ between its production and
detection, and if $\nu_m$ has mass $M_m$, then $A(\nu_m \mbox{ propagates}) =
exp(-i \frac{M_m^2}{2} \frac{L}{E})$ \cite{b2,b3,b4}.  Inserting these
amplitudes into \Eq{2}, we have 
\beq 
A(\nu_\ell \ra \nu_{\ell^\prime}) =
\sum_m U_{\ell m} e^{-i \frac{M_m^2}{2} \frac{L}{E}} U_{\ell^\prime m}^* .
\label{eq3}
\eeq
The probability $P(\nu_\ell \ra \nu_{\ell^\prime})$ for a neutrino of flavor
$\ell$ to oscillate in vacuum into one of flavor $\ell^\prime$ is then just
the square of this amplitude.

Note from \Eq{3} that oscillation from one flavor to another does require
neutrino mass. If all the masses $M_m$ vanish, then, since $U$ is unitary,
\Eq{3} implies that $A(\nu_\ell \ra \nu_{\ell^\prime}) = \sum_m U_{\ell
  m}U^*_{\ell^\prime m} = \delta_{\ell\ell\prime}$.

The quantum mechanics of neutrino oscillation is both fascinating and subtle
\cite{b5}. It continues to be explored \cite{b6}.

If only two flavors, $\ell$ and $\ell^\prime \neq \ell$, mix appreciably, $U$
takes the form
\beq
    U = \left( \begin{array}{cc}
        \cos \theta & e^{i\delta}\sin\theta \\
        -e^{-i\delta}\sin\theta & \cos\theta    \end{array} \right) \;.
\label{eq4}
\eeq
Here, $\theta$ is a mixing angle, and $\delta$ is a phase. Inserting \Eq{4}
into \Eq{3}, and supplying the factors of $\hbar$ and $c$ which we have been
omitting, we find that
\beq
    P(\nu_\ell \ra \nu_{\ell^\prime\neq\ell}) = \sin^2 2\theta \,
\sin^2 \left[ 1.27 \: \delta M^2 (\mbox{eV}^2) 
\frac{L(\mbox{km})}{E(\mbox{GeV})} \right]  .
\label{eq5}
\eeq
Here, $\delta M^2 \equiv M_2^2 - M_1^2$ is the splitting between the squared
masses of $\nu_1$ and $\nu_2$, the two neutrino mass eigenstates which make
up $\nu_\ell$ and $\nu_{\ell^\prime}$.

As we see from \Eq{5}, the oscillation probability $P(\nu_\ell \ra
\nu_{\ell\neq\ell^\prime})$ is not appreciable unless $\delta M^2
(\mbox{eV}^2) \, L(\mbox{km}) / E(\mbox{GeV})$ is of order unity or larger.
Thus, an experiment characterized by a given $L(\mbox{km}) / E(\mbox{GeV})$
will be able to probe neutrino mass splittings $\delta M^2 (\mbox{eV}^2)\,
\gsim [L(\mbox{km})/ E(\mbox{GeV})]^{-1}$. By making $L/E$ large enough, one
can probe very tiny values of $\delta M^2$. Herein lies the power of neutrino
oscillation searches.

Neutrino oscillation experiments are very commonly analyzed assuming that the
oscillation probability is correctly described by \Eq{5}. However, this
equation was derived assuming that only two neutrinos mix appreciably. If
three or more mix, \Eq{5} may not hold. Nevertheless, there are interesting
situations in which, even though more than two neutrinos are mixing, \Eq{5},
with the meaning of ``$\sin^2 2\theta$'' and ``$\delta M^2$'' slightly
modified, does hold. A second point concerns the possibility of CP violation.
Assuming that the observed CP-violating effects in K meson decays are due to
Standard Model physics, there must be CP-violating phase factors in the quark
analogue of the leptonic mixing matrix $U$. Thus, it is natural to expect
that there are CP-violating phases in $U$ as well. These phases could lead to
CP-violating effects such as $P(\bar{\nu}_\mu \ra \bar{\nu}_e) \neq P(\nu_\mu
\ra \nu_e)$ \cite{b7}. However, some neutrino oscillation measurements are
completely insensitive to any CP-violating phases in $U$.

An example which nicely illustrates both of these points is the scenario in
which there are three neutrino mass eigenstates, with two of them much
lighter than the third. That is, $M_{1,2} \ll M_3$. Imagine, now, that one
searches for $\optbar{\nu}_\ell \ra \optbar{\nu}_{\ell^\prime \neq \ell}$ in
an experiment with $L/E$ such that $M^2_3 (L/E) \sim 1$, so that $M^2_{1,2}
(L/E) \ll 1$. Then, in \Eq{3} for $A(\nu_\ell \ra \nu_{\ell^\prime})$, we
have $\exp (-iM^2_{1,2} L/2E) \cong 1$.  Thus, since the unitarity of U
requires that $U_{\ell 1}U^*_{\ell^\prime 1} + U_{\ell 2}U^*_{\ell^\prime 2}
= -U_{\ell 3}U^*_{\ell^\prime 3}$, \Eq{3} yields $A(\nu_\ell \ra
\nu_{\ell^\prime}) = U_{\ell 3}U^*_{\ell^\prime 3}[-1 + \exp (-iM^2_3
L/2E)]$. Squaring, and including the factors of $\hbar$ and $c$, we have
\beq
    P(\nu_\ell \ra \nu_{\ell^\prime\neq\ell}) = \left| 2 \; U_{\ell 3}
U_{\ell^\prime 3}\right|^2 \sin^2 \left[ 1.27 \: M^2_3 (\mbox{eV}^2) 
\frac{L(\mbox{km})}{E(\mbox{GeV})} \right]  .
\label{eq6}
\eeq
This oscillation probability has exactly the same mathematical form as the
two-neutrino formula, \Eq{5}, but with the role of ``$\sin^2 2 \theta$''
played by $\left| 2 U_{\ell 3}U_{\ell^\prime 3}\right|^2 $, and that of
``$\delta M^2$'' played by $M_3^2$.  Thus, even when more than two neutrinos
participate in the mixing, describing the data of an oscillation experiment
using the two-neutrino formula of \Eq{5} may still be a valid procedure.
Turning to the question of whether the experiment under discussion can see CP
violation, we recall that CPT invariance implies that one can obtain
$P(\bar{\nu}_\ell \ra \bar{\nu}_{\ell^\prime})$ from $P(\nu_\ell \ra
\nu_{\ell^\prime})$ by simply replacing $U$ by $U^*$ \cite{b7}. Since this
replacement does not affect \Eq{6} at all, we see that in our experiment,
$P(\bar{\nu}_\ell \ra \bar{\nu}_{\ell^\prime \neq \ell})$ is forced to be the
same as $P(\nu_\ell \ra \nu_{\ell^\prime \neq \ell})$, even if CP-violating
phases are present in $U$. Our experiment cannot uncover these phases \cite{b8}.

The presence of CP-violating phases could, however, be revealed by an experiment with
$L/E$ big enough so that even $M^2_{1,2} (L/E) \gsim 1$. To see this, we note
that when there are three flavors and three neutrino mass eigenstates, \Eq{3} and the
rule that $P(\bar{\nu}_\ell \ra
\bar{\nu}_{\ell^\prime})$ is the same as $P(\nu_\ell \ra \nu_{\ell^\prime})$
except that $U$ is replaced by $U^*$ imply that
\beq
    P(\nu_\ell \ra \nu_{\ell^\prime\neq\ell}) - P(\bar{\nu}_\ell 
\ra \bar{\nu}_{\ell^\prime \neq \ell}) = 4 \;\Im{\mathrm m} (U_{\ell 1} 
U^*_{\ell^\prime 1} U^*_{\ell 2} U_{\ell^\prime 2}) (s_{1 2} + s_{2 3} + s_{3 1}) \;.
\label{eq6a}
\eeq
Here, $s_{ij} \equiv \sin (\delta M^2_{ij} L/2E)$, where $\delta M^2_{ij}
\equiv M^2_i - M^2_j$. If $U$ contains CP-violating phase factors, the
quantity ${\Im\mathrm m} (U_{\ell 1} U^*_{\ell^\prime 1} U^*_{\ell 2} U_{\ell^\prime
 2})$ will be nonvanishing \cite{b9}. Thus, from \Eq{6a} we see that, in
an experiment in which $\delta M^2_{ij} L/2E$ is not tiny compared to unity
for any $i \neq j$, the CP-violating asymmetry $P(\nu_\ell \ra
\nu_{\ell^\prime\neq\ell}) -
P(\bar{\nu}_\ell \ra \bar{\nu}_{\ell^\prime \neq \ell}) \equiv
A_{\mathrm CP}(\ell\ell^\prime)$ will be nonvanishing
as well. Interestingly, the unitarity of $U$ is readily shown to imply that
$A_{\mathrm CP}(\ell\ell^\prime)$, as given by \Eq{6a}, has the same value for
$\ell\ell^\prime = e\mu,\;\mu\tau$, or $\tau e$. Furthermore, \Eq{6a} obviously
implies that $A_{\mathrm CP}(\ell^\prime\ell) = -A_{\mathrm CP}(\ell\ell^\prime)$.
Thus, when there are only three neutrinos, all the CP asymmetries $A_{\mathrm
CP}(\ell\ell^\prime)$ are equal to within a sign. However, if, say, $\delta M^2_{12}
L/2E \ll 1$, then $s_{1 2} + s_{2 3} + s_{3 1} \cong 0$, and all the asymmetries
$A_{\mathrm CP}(\ell\ell^\prime)$ disappear. This is the situation we
encountered in the example we just discussed. Obviously, the CP asymmetries also
disappear if $\delta M^2_{23} L/2E$ or $\delta M^2_{31} L/2E$
is much smaller than unity.

In Section~\ref{sec2}, we saw that the see-saw mechanism predicts the existence
of {\em very heavy} neutral leptons in addition to the observed three light
neutrinos. It may be that there are also {\em light} neutrinos in addition to
the observed $\nu_e$, $\nu_\mu$, and $\nu_\tau$. It is known that, should
there be any such additional light neutrinos, the weak boson $Z$ does not
couple to them \cite{b10}. This means that these additional neutrinos do not
participate in the normal weak interactions of the SM.  That is, they are
what are called ``sterile'' neutrinos.

If there do exist light sterile neutrinos, then any of the weak-interaction
``active'' neutrinos, $\nu_e$, $\nu_\mu$, or $\nu_\tau$, can oscillate, not
only into another active neutrino, but into a sterile neutrino as well. The
mathematics of neutrino oscillation when sterile neutrinos are involved is
the same as when they are not. One simply increases the number of neutrino
flavors, the number of neutrino mass eigenstates, and the consequent size of
the mixing matrix $U$, to accommodate the new neutrinos of sterile flavor.

Owing to the special nature of the SM weak currents which produce and absorb
neutrinos, many neutrino processes are insensitive to whether the neutrinos
are Dirac or Majorana particles \cite{b11}. By and large, neutrino oscillation
is insensitive as well. However, if the neutrino mass eigenstates are
Majorana particles, then in addition to the oscillation of one neutrino
flavor into another, which we have been discussing, we can also have the
oscillation of a neutrino into an antineutrino. That is, a neutrino which in
interaction with matter produces a negatively-charged lepton $\ell^-$
(conventionally called a ``lepton'') can become one which produces a
positively-charged lepton $\ell^+$ (conventionally called an ``antilepton'').
There are several different varieties of this $\nu \ra \bar{\nu}$
oscillation. To understand these varieties, let us recall that, irrespective
of whether neutrinos are of Majorana or Dirac character, neutrino
interactions obey certain helicity rules \cite{b11}. When the SM weak
interactions produce an $\ell^+ (\ell^-)$, the accompanying neutrino or
antineutrino almost (but not quite) certainly has negative (positive)
helicity. The amplitude for it to have positive (negative) helicity is
suppressed, and is only of order $M/E$, where $M$ and $E$ are, respectively,
the mass and energy of the neutrino. Correspondingly, when the SM weak
interactions absorb a neutrino or antineutrino to produce an $\ell^-
(\ell^+)$, the absorbed particle must have negative (positive) helicity. The
amplitude for it to be absorbed and produce a lepton of the desired charge if
it has the wrong helicity is, again, only of order $M/E$. However, these
suppressed amplitudes are not zero.

When neutrinos are Dirac particles, there is a conserved lepton number $L$,
with $L(\ell^-) = L(\nu) = -L(\ell^+) = -L(\bar{\nu}) = +1$. Conservation of
$L$ prevents a neutrino from being born in association with an $\ell^+$ and
then interacting to make a second $\ell^+$ instead of an $\ell^-$. But when
neutrinos are Majorana particles, no such conservation law stands in the way.
The neutrino need only have the proper helicity. Now, as has been stated,
when a ``neutrino'' is born in association with an $\ell^+$, this neutrino,
while dominantly of negative helicity, has a small component, of order $M/E$,
with positive helicity. This small component has the right helicity to make
an $\ell^+$ when interacting with matter.  Since a particle whose interaction
with matter yields an $\ell^+$ is customarily called an ``antineutrino'', we
would say that the particle born as a $\nu$ became a $\bar{\nu}$. However, it
is probably not too accurate to say the $\nu$ ``oscillated'' into a
$\bar{\nu}$, since a small piece of it had the correct helicity to act as a
$\bar{\nu}$ right from the moment of birth, without having to travel any
distance at all. This is the first variety of $\nu \ra \bar{\nu}$
``oscillation''.

A quite different kind of $\nu \leftrightarrow \bar{\nu}$ oscillation is the
neutrino analogue of $B^o \leftrightarrow \bar{B^o}$ meson oscillation. Like
the latter, which has been observed \cite{b12}, this would be a genuine
oscillation with distance. It is sometimes referred to as 2nd class
oscillation \cite{b13}.  To illustrate how this type of oscillation works, let
us neglect intergenerational mixing, and imagine that the electron neutrino
with negative helicity, $\nu_e ^\downarrow$, and its charge conjugate, the
electron antineutrino with negative helicity $\bar{\nu}_e ^\downarrow$, are
orthogonal linear combinations of two underlying Majorana mass eigenstates.
Here the arrows indicate the helicity.  This situation is analogous to the
one in the $B^o - \bar{B^o}$ system, where $B^o$ and $\bar{B^o}$ are
orthogonal linear combinations of two mass eigenstates. Just as a $B$ meson
born as a $B^o$ then oscillates between $B^o$ and $\bar{B^o}$ states, so a
neutrino born as a $\nu_e ^\downarrow$ would oscillate between $\nu_e
^\downarrow$ and $\bar{\nu}_e ^\downarrow$. Note that the latter oscillation
necessarily connects two states of the same helicity, since the angular
momentum of a neutrino in vacuum is conserved.  Now, a $\bar{B^o}$
participates in SM interactions, but a $\bar{\nu}_e ^\downarrow$ does not,
being of the wrong helicity to do so. That is, this brand of ``$\nu
\leftrightarrow \bar{\nu}$ oscillation'' turns an active neutrino, $\nu_e
^\downarrow$, into a sterile one, $\bar{\nu}_e ^\downarrow$. In particular,
this sterile neutrino does not interact to produce a positron, as a
$\bar{\nu}_e ^\uparrow$ would.  In an experiment starting with a $\nu_e$
beam, $\nu_e ^\downarrow \ra \bar{\nu}_e ^\downarrow$ would be marked by a
disappearance of some (possibly large) fraction of the original $\nu_e$ flux,
but without the appearance of a corresponding flux of new particles capable
of interacting to produce a charged lepton of any charge or flavor. Of
course, with probability $\sim (M/E)^2$, a $\bar{\nu}_e ^\downarrow$ can
indeed produce an $e^+$, but this is a very small effect.
Neutrino-antineutrino oscillation is a potentially very interesting effect, but
estimates of its experimental visibility are discouraging \cite{b14}.

\subsection{Flavor Transitions in Matter}   \label{ssec31}

When a neutrino propagates through matter, rather than through a vacuum, its
ability to change its flavor can be greatly enhanced by its interaction with
the surrounding medium. This phenomenon, known as the
Mikheyev-Smirnov-Wolfenstein (MSW) effect \cite{b15}, results from weak
forward coherent scattering from electrons, and may play a very
important role in the behavior of the solar neutrinos (see Section~\ref{sec42}).
A solar neutrino is created as a $\nu_e$ in the core of the sun by nuclear
processes.  The solar neutrino detectors presently operating on earth can
detect a $\nu_e$, but are completely, or at least largely, insensitive to
neutrinos of other flavors. However, as the solar neutrino born as a $\nu_e$
journeys outward from the solar core through solar material, the MSW effect
can potentially convert it into a neutrino $\nu_x$ of muon, tau, or even
sterile flavor. It would then be undetectable in the present detectors. Thus,
the solar neutrino flux observed by these detectors would be less than that
predicted when one does not take this MSW flavor conversion into account.

Like neutrino oscillation in vacuum, MSW flavor conversion requires neutrino
mass. Without going through the physics of the MSW effect in detail, let us
recall enough of it to see how much mass it requires.

The MSW effect converts a $\nu_e$ into a $\nu_x$ ($=\nu_\mu, \nu_\tau$, or
$\nu_{\mbox{\scriptsize{sterile}}}$) as a result of a crossing, somewhere in
the sun, of the $\nu_e$ and $\nu_x$ energy levels. Neglecting neutrino
mixing, the total energy of a $\nu_e$ of mass $M_{\nu_e}$ and momentum $p$ in
the sun is $\sqrt{p^2 + M_{\nu_e}^2} + \sqrt{2}G_F N_e$. Here, $G_F$ is the
Fermi coupling constant, and $N_e$ is the electron density at the location of
the neutrino. The term $\sqrt{p^2 + M_{\nu_e}^2}$ is the energy the $\nu_e$
would have in empty space, and the term $\sqrt{2}G_F N_e$ is an interaction
energy arising from the W-boson exchange interaction between the $\nu_e$ and
solar electrons. We have neglected a small Z-boson exchange interaction.
Now, a $\nu_x$ ($=\nu_\mu, \nu_\tau$, or $\nu_{\mbox{\scriptsize{sterile}}}$)
does not have a W-boson exchange interaction with electrons. Thus, neglecting
the small Z-boson exchange interaction once again, the total energy of a
$\nu_x$ of mass $M_{\nu_x}$ and the same momentum $p$ as before is simply
$\sqrt{p^2 + M_{\nu_x}^2}$. The condition that the $\nu_e$ and $\nu_x$ energy
levels cross is then
\beq
    \sqrt{p^2 + M_{\nu_x}^2} \cong \sqrt{p^2 + M_{\nu_e}^2} + \sqrt{2}G_F N_e \; .
\label{eq7}
\eeq
Taking for $N_e$ a characteristic solar electron density, $\sim 10^{26}$/cc,
and for $p$ a characteristic solar neutrino momentum, $\sim 1$ MeV, we find
that if this level-crossing condition is to be satisfied, we must have
\beq
    M_{\nu_x}^2 - M_{\nu_e}^2 \equiv \delta M_{\nu_x\nu_e}^2 \sim 10^{-5} \mbox{eV}^2
\; .
\label{eq8}
\eeq
This is the amount of mass (splitting) required by the MSW effect. Note, in
particular, that the MSW effect does require nonzero neutrino mass.
\footnote{Of course, if there should exist flavor-changing neutrino
  interactions such as $\nu_e + \mbox{nucleon} \;\ra \;\nu_\mu +
  \mbox{nucleon}$, then a neutrino could change flavor in matter even in the
  absence of neutrino mass. However, we assume that, as in the SM, there are
  no neutrino-flavor-changing interactions. Then flavor conversion requires
  neutrino mass.}


\subsection{Types of Neutrino Oscillation Experiments}\label{sect:types}

Neutrino flavor oscillation experiments fall crudely into two categories:
disappearance experiments, and appearance experiments.  
If an experimenter creates a beam of neutrinos of lepton flavor $\ell$, then 
in a disappearance experiment, he or she will search for the absence of some of
the $\nu_\ell$ flux known to be present in the initial beam.  In an
appearance experiment, he or she will seek the appearance of neutrinos
$\nu_{\ell^\prime \ne \ell}$.  The appearance of a lepton flavor not present
in the original beam is often a more convincing signal of oscillation than is
the disappearance of some fraction of a neutrino beam whose total flux may be
uncertain. However, it may be that in some cases neutrino oscillations
convert active neutrinos into sterile ones. To discover this, an experimenter
would have to demonstrate both the disappearance of the original active
flavor and the non-appearance of any new active flavor in its stead.

Experiments seeking neutrino flavor oscillations typically use
charged-current neutrino interactions ($\nu_\ell+A\to\ell+B$) to tag the
flavor of the observed neutrinos.  Experiments with low neutrino energy,
which are desirable to increase $L/E$ for the purposes of probing low $\delta
M^2$, are often unable to use this technique because of the lack of energy
necessary to produce the massive charged lepton.  Observations of
neutral-current neutrino interactions ($\nu_\ell+A \to \nu_\ell+B$) are
important probes of the sterile neutrino hypothesis, since neutral-current
interactions would be absent for sterile neutrinos but present for active
neutrinos even when the related charged-current interactions are suppressed
by the charged lepton mass.

A particular case of interest occurs when backgrounds and intensities are
low, such as in many ``long-baseline'' (large $L$) experiments.  Appearance
experiments have sensitivities which scale with $N\times P$, where $N$ is the
number of neutrinos expected in the detector and $P$ is the probability of
flavor oscillation.  However, disappearance experiments have sensitivities
which scale as $P\sqrt{N/(1-P)}$, and are therefore often less powerful at low
mixing angle in the case of transition from $\nu_{\rm active}\to\nu^{'}_{\rm
active}$.

Interestingly, CP violation in neutrino oscillation could never be seen in a
disappearance experiment. Such an experiment simply measures the probability
$P(\nu_\ell \ra \nu_\ell)$ that a neutrino retains its original flavor.
Violation of CP would be signalled by $P(\bar{\nu}_\ell \ra \bar{\nu}_\ell)
\neq P(\nu_\ell \ra \nu_\ell)$. However, one can show that CPT invariance
demands that $P(\bar{\nu_\ell} \ra \overline{\nu}_\ell) = P(\nu_\ell \ra
\nu_\ell)$.

In order to establish oscillation as the correct explanation of an observed
phenomenon, it is desirable to observe the characteristic variation in
neutrino beam composition with $L/E$.  This may appear as a spectral
distortion of a wideband source of neutrinos or as a change in observed
appearance rates as $L/E$ varies in a narrowband source.


\section{THE OBSERVED EVIDENCE AND HINTS OF NEUTRINO OSCILLATION}

\subsection{Atmospheric Neutrinos}\label{sect:atmos}

Atmospheric 
neutrinos are produced by interactions of cosmic ray protons with air
nuclei in the earth's atmosphere via the reaction chain
\begin{eqnarray}
p + 
N & \ra & \pi^{\pm} + X \\
\pi^{\pm} & \ra & \mu^{\pm} + \numu (\numubar) \\
\mu^{\pm} & \ra & e^{\pm} + \nue (\nuebar) + \numubar (\numu). 
\end{eqnarray}
From simple counting the flavor ratio resulting from this chain is 
\begin{eqnarray}
\Rmue = \frac{N(\numu + \numubar)}{N(\nue + \nuebar)} \sim 2. \label{eq:ratio}
\end{eqnarray}
A more sophisticated calculation changes this ratio somewhat, but
calculations from different groups\cite{ref:atnucalc} are consistent at the
5\% level.  The absolute fluxes are believed to be known at the 20\% level.
This makes neutrinos produced in the atmosphere a potentially good
place to look for oscillation effects: the ratio between species is
reasonably well known even though the absolute fluxes are not.

Generally, an oscillation effect would show up as a variation of the above
ratio from the calculated value of approximately $2$.  
Detailed calculations of flavor ratios are used to account for 
detector effects, such as energy thresholds and acceptances which vary from
experiment to experiment and deviations from $R_{\mu/e}=2$ at
higher neutrino energies.
Conventionally, experimenters report the ratio between the measured value 
of the quotient in 
\Eq{:ratio}, and its predicted value based on a Monte Carlo simulation:
\begin{equation}
\Rdatamc= \frac{\left[ N(\numu + \numubar)/N(\nue + \nuebar) \right] _{\rm Data}}
{\left[ N(\numu + \numubar)/N(\nue + \nuebar) \right]_{\rm MC} } \; .
\label{eq:ratio2}
\end{equation}
This ``ratio of ratios'' should be unity if there are no neutrino oscillations and if
detector
effects and backgrounds 
are correctly modeled.  If $\numu$ flavor oscillations occur,
$\Rmue$ would be less than two and $\Rdatamc$ would be less than one.

The zenith angle distribution of neutrino events of a given flavor
provides a second, more sensitive,
method of detecting oscillations which is less dependent on theoretical input.
Neutrinos produced in the atmosphere just above the detector will have a 
shorter flight path (20 km) 
than neutrinos produced on the opposite side of the 
earth, which travel up to 13,000 km before detection.  Thus, the oscillation 
baseline varies with the zenith angle $\theta_Z$.  
In the absence of oscillations, the
zenith angle distribution will be up-down symmetric to a good approximation, and any 
deviation from this symmetry will indicate a change in the neutrino flux during
passage through the earth.  As this point may not be immediately clear, we give the
argument in detail below.

\begin{figure}    
    \hspace*{0.2\textwidth}\epsfxsize=0.6\textwidth\epsfbox{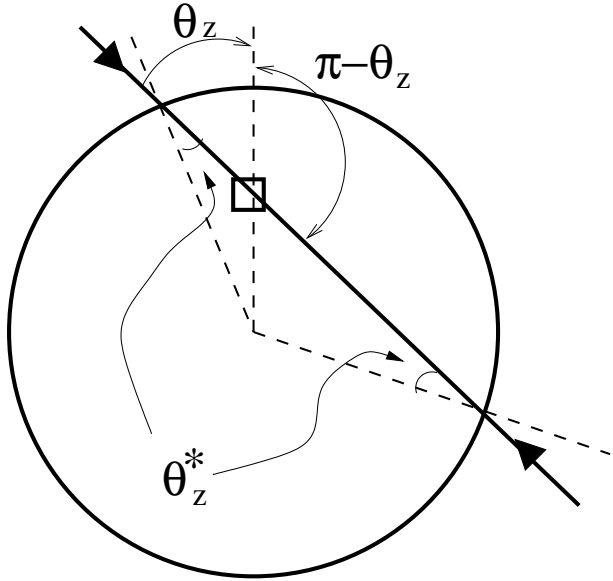}
    \caption{Atmospheric neutrinos incident on an underground detector from
      zenith angles $\theta_Z$ and $\pi - \theta_Z$. 
The large circle is the
      earth, and the black square the detector. The various dashed lines intersect       
at the center of the earth.
      The neutrinos travel as shown by the arrows along the solid line from
      their points of origin just above the earth's surface to the detector. }
      \label{f1}
\end{figure}

Suppose that neutrinos do not oscillate. More
generally, suppose that the number of neutrinos of a given flavor does not
change for any reason as the neutrinos travel from their points of birth to
the detector. Then, at neutrino energies above a few GeV, where geomagnetic
effects can be neglected, the detected flux of neutrinos of a given flavor coming from
zenith angle $\theta_Z$ must be equal to that coming from angle $\pi - \theta_Z$. To
see why, consider Figure~\ref{f1}.

Measurements have found that above a few GeV the cosmic ray flux which
produces the atmospheric neutrinos is isotropic \cite{b17}. Thus, neutrinos
are being created in equal numbers per unit area and time everywhere around
the earth in Figure~\ref{f1}. In addition, the angular distribution of the
produced neutrinos, relative to the local downward-pointing vertical, is the
same around the world. We shall call this angular distribution $D(\theta_Z^*)$,
where $\theta_Z^*$ is the angle between the momentum of the neutrino and the
downward vertical. Consider, now, the downward flux of neutrinos incident on
the detector from a small solid angle $\Delta \Omega$ at zenith angle
$\theta_Z$. Let us compare this to the corresponding upward flux from the same
size solid angle at $\pi - \theta_Z$. Note from Figure~\ref{f1} that the angle
$\theta_Z^*$ between the momentum the neutrinos must have to reach the detector
and the downward vertical at the neutrino production point is the same for
the neutrinos coming from $\pi - \theta_Z$ as for those coming from $\theta_Z$.
Thus, $D(\theta_Z^*)$ has the same value for both. For a $\Delta \Omega$ of
given size, the surface area of the region of atmosphere that contributes
neutrinos is proportional to $\ell^2$, where $\ell$ is the distance from the
detector to the contributing region. This effect would appear to favor the
upward-going neutrinos from the distant source at $\pi - \theta_Z$ over the
downward-going ones from the nearby source at $\theta_Z$. However, for the
neutrinos coming from a tiny area $da$ of fixed size within the source
region, the solid angle subtended by the detector falls off as $1/\ell^2$.
Thus, when one integrates over the contributing source region, the factors of
$\ell^2$ cancel out, and neither the upward- nor downward-going neutrinos are
favored. Finally, since the source region of atmosphere is not perpendicular
to the line connecting it to the detector, for given $\Delta \Omega$ the size
of this region is enhanced by a factor of $\sec \theta_Z^*$. However, since
$\theta_Z^*$ is the same for the neutrinos from $\theta_Z$ and from $\pi -
\theta_Z$, this factor, like all the others, does not favor either of these
neutrinos.

The conclusion is that, per unit solid angle, the fluxes of neutrinos of a
given flavor incident on the detector from directions $\theta_Z$ and $\pi -
\theta_Z$ must be equal. If they are not, then the number of neutrinos of this
flavor must change as the neutrinos journey from their points of origin to
the detector. Most likely, the change is due to vacuum neutrino oscillation,
although one must be careful to consider other possibilities, such as the
decay of neutrinos of one flavor into those of
another\cite{ref:nudk}. 

\subsubsection{Experimental Results}

The first generation detectors which studied atmospheric neutrinos,
NUSEX\cite{ref:nusex}, Soudan\cite{ref:soudan}, IMB\cite{ref:IMB},
Frejus\cite{ref:frejus} and Kamiokande\cite{ref:kamii}, 
were originally designed
to search for proton decay.  All are deep underground to reduce backgrounds
from cosmic rays, all are very massive in order to be sensitive to proton
lifetimes of order 10$^{32}$~yr and designed in order to minimize backgrounds
from radioactive substances.  These characteristics make them ideal for the
detection of both solar and atmospheric neutrinos: the neutrino flux from 
cosmic ray interactions in the atmosphere is expected to be
$\sim 10^9$ m$^{-2}$-sr$^{-1}$-($100$~MeV)$^{-1}$-yr$^{-1}$, at 
1~GeV, so there are $\sim 50$ reactions per
kiloton of target mass per year.

For the second generation of detectors, Soudan II\cite{ref:soudan},
Kamiokande III\cite{ref:kam2} and IMB-3\cite{ref:IMB}, the key improvement was
lowering of the energy threshold of the detector trigger.  Generally, this
required the use of lower background materials and noise reduction in the
detector systems as well as the ability to handle higher data 
rates.  SuperKamiokande\cite{ref:superk} represents the third
generation of proton decay experiments with neutrino detection placed on an
equal footing with proton decay.  SuperKamiokande's larger fiducial volume not
only means a higher counting rate from neutrino interactions, but the ability 
to fully contain a larger fraction of the final state products from the
neutrino interactions.

The large underground detectors are of two types: water \v{C}erenkov (IMB-3,
Kamiokande and SuperKamiokande), which use water as the target and measure the
\v{C}erenkov light cone to determine a particle's velocity and direction, 
and tracking detectors
(Soudan, Frejus, NUSEX and MACRO), which use dense material as a target and 
sampling detectors to identify the final state lepton.

The most sensitive result comes from SuperKamiokande, the latest of the water
\v{C}erenkov detectors, and we describe it in some detail. A particle moving through a
medium of index of refraction $n$ with a velocity greater than $c/n$ will radiate
photons into a forward cone of opening angle $\cos\theta_c =
1/n\beta$. For ultra-relativistic particles in water, the \v{C}erenkov
angle is $\theta_c=41.7^\circ$, and a charged particle moving
through water will generate around nine photo-electrons per MeV of
deposited energy.  The arrival time of the photons at the photon
detectors allows the reconstruction of the \v{C}erenkov cone and hence
the velocity and direction of the charged particle.  The 200 ps
timing resolution of the photon detectors gives a vertex resolution
of 30 cm.  The total path length gives the total energy with a
resolution of $2.5\%/\sqrt{E\mathrm{(GeV)}} \pm 0.5\%$ for electrons
and 3\% for muons.

Two key ingredients make measurements of atmospheric neutrinos possible in
water \v{C}erenkov detectors.  First, even at a depth of  1000 meters 
(equivalent to 3000 meters of water), the cosmic
muon flux is $10^8 \mu$/year in SuperKamiokande compared to the 
observed change in the flux (see below), which is around 150 events for a year
of running.
Requiring tracks to originate at least 2 m from the surface of the detector
reduces the accidental flux by a factor of 10$^9$ for the fiducial mass of 22
kilotons of this detector \cite{ref:superkrates}.  
Second, the very large number of \v{C}erenkov photons
allows
discrimination between electrons and muons based on the larger multiple
scattering of the electrons owing to their low mass\cite{ref:skpid}.  The
higher multiple scattering of electrons leads to a fuzzier \v{C}erenkov ring
image; measuring the sharpness of this image allows discrimination between
electrons and muons with less than 1\% confusion.  

Fully contained events are classified according to their visible energy
(``sub-GeV'' for $E_{\rm vis} < 1.33 $GeV, ``multi-GeV'' for
$E_{\rm vis}>1.33$ GeV) and the number of reconstructed \v{C}erenkov rings
(``single-ring'' for one reconstructed ring, ``multi-ring'' for two or
more).  Single ring events are further classified based on whether the
ring originates from an electron (``e-like'') or a muon (``$\mu$-like'').
Identification of the final state lepton is not possible for multi-ring events.  
Events in which the final state lepton (usually a muon) exits the fiducial region are
considered partially-contained (``PC'') and treated separately
\cite{ref:skmulti} from fully contained (``FC'') events.

Turning first to the total rate, two independent published
 analyses (``A'' and ``B'') \cite{ref:superkrates}
for the sub-GeV events, and a separate analysis for multi-GeV events 
\cite{ref:superk}, 
give 
\begin{eqnarray}
        & = 0.61 \pm 0.03 ({\rm stat.}) \pm 0.05 ({\rm syst.}),
& \mathrm{sub-Gev, A}\\
\Rdatamc & = 0.65 \pm 0.03 ({\rm stat.}) \pm 0.05 ({\rm syst.}),
& \mathrm{sub-GeV, B}\\
        & = 0.65 \pm 0.05 ({\rm stat.}) \pm 0.08 ({\rm syst.}),
& \mathrm{multi-GeV}
\end{eqnarray} 
The most recent results~\cite{ref:skmess99} combining FC and PC muons are
\begin{eqnarray}
\Rdatamc & = & 0.67 \pm 0.02 \pm 0.05 \;{\rm , sub-GeV}\\
         & = & 0.66 \pm 0.04 \pm 0.08 \;{\rm , multi-GeV}
\end{eqnarray}

The dominant uncertainty in each analysis comes from the calculation of the
$\nu_{\mu}/\nu_e$ flux ratio (5\%) and the neutrino charged current and
neutral current cross sections (3-3.5\% each).  The shape of the momentum
distribution of the final state lepton in single-ring events agrees with the
Monte Carlo expectation, hence $\Rdatamc$ appears to be independent of
energy.

\begin{table}
\begin{center}
\begin{tabular}{|l r r ||l r r|}
\hline
\multicolumn{3}{| c ||}{Rate Analysis} & \multicolumn{3}{| c |}{Zenith Angle
Analysis} \\
\hline
Analysis A  & Data & Monte Carlo & sub-GeV & Data  & Monte Carlo \\     
\hline
single-ring & 1883 & 2030.5        & single-ring & 2389 & 2622.6 \\
e-like      & 983 & 821.2       & e-like      & 1231 & 1049.1 \\
$\mu$-like  & 900 & 1218.3      & $\mu$-like  & 1158 & 1573.6 \\
multi-ring  & 784  & 759.2       & multi-ring  &  911 & 980.7    \\
\hline
total       & 2579 & 2789.7      & total       & 3300 & 3603.3 \\
\multicolumn{3}{| c ||}{$R=0.61 \pm 0.03 \pm 0.05$} &
\multicolumn{3}{| c |}{} \\
\hline
\hline
Analysis B  & Data & Monte Carlo & multi-GeV   & Data   & Monte Carlo \\
\hline
single-ring & 2008 & 2185        & single-ring & 520    & 531.7  \\
e-like      & 967  & 812.1       & e-like      & 290    & 236    \\
$\mu$-like  & 1041  & 1364.8      & $\mu$-like  & 230    & 295.7  \\
multi-ring  & 642  & 631.3       & multi-ring  & 533    & 560.1  \\
\hline     
total       & 2650 & 2817.2      & total       & 1053   & 1091.8  \\
\multicolumn{3}{| c ||}{$R=0.65 \pm 0.03 \pm 0.05 $} & 
\multicolumn{3}{| c |}{} \\
\hline
            &                    & & PC          & 301    & 371.6   \\
\hline
\end{tabular}
\caption{\label{ta:skdata} Numbers of events used in rate and zenith angle
analyses.  Left panel are number of events of various types used in two separate
rate analyses from \cite{ref:superkrates}.  Right panel are the event
categories used in the zenith angle analysis 
\cite{ref:superk}.}
\end{center}
\end{table}

Combining the statistical and systematic errors in quadrature gives
\begin{eqnarray}
\Rdatamc & < & 0.78~~{\rm at}~95\%~{\rm c.l.}
\end{eqnarray}
The average neutrino flight path is 6000 km, which gives
\begin{eqnarray}
\delta M^2 & > & 2 \times 10^{-4} {\rm eV}^2~~{\rm at}~95\%~{\rm c.l.}
\end{eqnarray}

Since the number of oscillations depends on neutrino flight
distance, $\Rdatamc$ should vary with the zenith angle. So too should the $\nu_\mu$
event rate and/or the $\nu_e$ event rate out of which $\Rdatamc$ is formed. The
direction of the final state lepton approximates the neutrino flight direction
within 55$^{\mathrm{o}}$ at $p_l$=400 MeV/c and 20$^{\mathrm{o}}$ at 
$p_l$=1 GeV/c. The path length is a very strong function of zenith angle: the average
flight length for neutrinos produced above the horizon
($\cos\theta_Z>0$) is 74 km while the path length for those produced
below the horizon ($\cos\theta_Z<0$) is 6500 km. 
As previously argued, in the absence of oscillation 
the detected flux of multi-GeV neutrinos is expected to be up-down symmetric, and
detailed calculations indicate that any asymmetry should be less than a few percent.

Let $A = (U - D) / (U + D)$ be the asymmetry between the number of 
upward-going ($\cos\theta_Z < -0.2$) muons, $U$, and the number of 
downward-going ($\cos \theta_Z > 0.2$) ones, $D$. 
From the data in Table~\ref{ta:skdata}, published in \cite{ref:superk}, 
\begin{eqnarray}
A & = & -0.296 \pm 0.048 \pm 0.01~~{\rm multi-GeV,}~{\rm FC+PC }~\mu 
\end{eqnarray}
The updated result is \cite{ref:skmess99}
\begin{eqnarray}
A & = & -0.311 \pm 0.043 \pm 0.01~~{\rm multi-GeV,}~~{\rm FC+PC}~\mu. 
\end{eqnarray}
In both cases, the asymmetry for electron events is consistent with
zero.

The nonzero asymmetry for neutrino induced muon events clearly indicates
something is happening: muon neutrinos are being lost as they transit
the earth. The detailed zenith angle distribution is well fit by the 
hypothesis that the disappearing muon neutrinos are undergoing
two-flavor oscillation into tau or perhaps sterile 
neutrinos.\footnote{Reactor neutrino oscillation experiments (see Section 6.1) probe
the $1.5 \times 10^{-3} < \delta M^2 < 6 \times 10^{-3}$ eV$^2$ region using
$\nuebar$ disappearance, which is not observed.  By CPT conservation,
$\nuebar\not\to\numubar~\Rightarrow~\numu\not\to\nue$, and therefore the
reactor experiments rule out $\numu \ra
\nue$ oscillations as the origin of the deficit of atmospheric muon neutrinos.} 
Assuming $\nu_\mu \ra \nu_\tau$, the favored values of the oscillation parameters are
\begin{equation}
5 \times 10^{-4}~<~\delta M^2~<~6\times 10^{-3} {\rm eV}^2~~{\rm  and }~\sin^2
2\theta >0.86
\end{equation}
from the published analysis. For the updated results, with maximum mixing, which is
favored, the 95\% c.l. interval for $\delta M^2$ is  
\begin{equation}
1.5 \times 10^{-3}~<~\delta M^2~<~6 \times 10^{-3}\; {\rm eV}^2 \;.
\end{equation}
The allowed oscillation region is shown in Figure~\ref{fi:excl}.

Neutrino decay has been considered as a mechanism to account for the loss of
muon neutrinos traveling through the earth \cite{ref:nudk}.  Although
neutrino decay can account for the missing $\numu$ flux, neutrino decay in
vacuum does not account for the energy dependence of the zenith angle
distribution observed experimentally.  Including the effects of neutrino
decay in matter improves the agreement somewhat, but leads to an
underestimate on the number of upward going muon events
\cite{ref:nudkmat}. 
  

\subsubsection{Comments on the Results}

At this point, it is useful to consider the numbers of events 
involved.  Assuming the flux of electron neutrinos remains unchanged 
during transit through the earth, the number of observed e-like
single ring events divided by the number expected gives a flux
normalization of 1.17; the observed flux is 17\% larger than predicted.
Normalizing the number of expected single ring mu-like events by this
factor gives an observed deficit of 550 mu-like events.  Normalizing the 
number of multi-ring events in the same way gives a deficit of about
100 events, which is consistent with the number of $\numu$ induced
charged current events lost if $\Rdatamc$=0.6. 
For the zenith angle 
asymmetry, if $A=-0.3$, the number of upward ($\cos\theta_Z< -0.2$) 
mu-like single ring events, $U$, differs from the number of downward 
($\cos\theta_Z>0.2$) events, $D$, by $U-D \sim 110$ 
for the multi-GeV single
ring sample and $\sim 150$ for the sub-GeV single ring sample. Thus, the conclusion
that neutrinos oscillate rests on the deficit of a few hundred events in the flux
ratio measurement, and about one hundred in the zenith angle measurement.

\begin{figure}
\epsfxsize=\textwidth
\epsfbox[40 280 430 560]{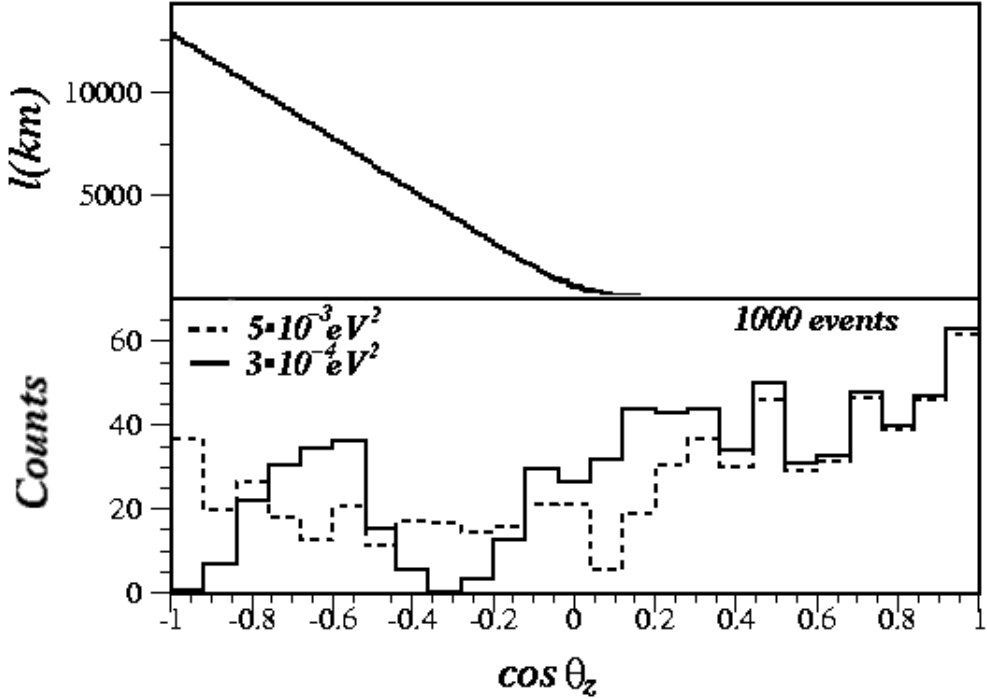}
\caption{\label{fig:sk} Zenith angle measurement.  a) shows relation
  between $\cos\theta_Z$ and distance between production point and detector,
  b) shows the number of counts as a function of $\cos\theta_Z$ for selected
  $\dmsq$ and maximal mixing with no angular smearing assuming $1$~GeV
  atmospheric neutrinos.  At $1$~GeV, the SuperKamiokande angular resolution
  is $20^{\mathrm{o}}$.}
\end{figure}

Taking a critical view, several ideas come to mind.  Most obviously,
misidentification of cosmic rays could increase the number of downward
going muons which, when combined with the 20\% absolute normalization
uncertainty, could give the observed asymmetry.  
The angular resolution on the incoming neutrino direction (which
arises primarily owing to the approximation of the incoming neutrino
direction by the direction of the recoil lepton) largely washes out
the oscillation signal and the measurement really amounts to a two bin
($\cos\theta_Z>0.2$ and $\cos\theta_Z<-0.2$) measurement, as clarified by
Figure~\ref{fig:sk}.  The effect could then be caused by an
underestimate of the amount of background from downward going cosmic ray muons.
Given that 10$^8$
cosmic ray muons traverse the detector each year, even a very small
error in the veto efficiency would lead to a large asymmetry.
However, Superkamiokande's outer detector acts independently of the
inner detector, allowing independent veto calibration of the two
systems.  In addition, no systematic variation of $A$ is observed
when the vertex cut is relaxed to allow events closer to the surface
of the inner detector \cite{ref:superkrates}.  
The claim of complete absence of cosmic ray background in the final sample seems
compelling.

Geomagnetic effects could also play a large role, especially for
neutrinos with $E_{\nu}<$ 1 GeV which would originate from lower
energy primary particles.  
The earth's magnetic field bends cosmic ray primaries below a few GeV
away from the earth
near the equator where the component of the magnetic field perpendicular
to the direction of travel of the cosmic rays is largest ($\sim 0.3$G).
Near the magnetic poles, the perpendicular component goes to zero, allowing
cosmic rays of all energies to interact in the atmosphere.
Downward going neutrinos originate from
interactions which take place at latitude 40$^o$N, while upward going
neutrinos originate from essentially all latitudes.  One would suspect
this would lead to the downward going neutrinos having a harder energy
spectrum owing to the harder spectrum of the primaries.  Calculation
reveals the size of the effect has only a 1-2\% impact on
$A$ \cite{ref:magcalc} for the multi-GeV sample, 
much too small to account for the observed asymmetry.
In addition, SuperKamiokande has measured the azimuthal (or ``East-West'')
distribution of neutrino events from near the horizon\cite{ref:skew} and observes
the geomagnetic asymmetry predicted by \cite{ref:magcalc}.

Atmospheric neutrinos were extensively studied by the other large underground
experiments.  Almost all observe a deficit of $\mu$-like events leading to a
low value for $\Rdatamc$, Table~\ref{ta:atmother}.  Particularly significant
in these results is the confirmation of this anomaly in both major types of
detectors, water \v{C}erenkov (Kamiokande, IMB-3 and Superkamiokande) and dense
sampling detectors (Soudan-2).

\begin{table}
\begin{center}
\begin{tabular}{|l c l|}
\hline
Experiment & Exposure & $\Rdatamc$ \\
           & (kt-y) & \\
\hline
Soudan-2\cite{ref:soudan2} 
& 3.9 & $0.66 \pm 0.11^{+0.05}_{-0.06}$ \\
IMB-3\cite{ref:imb3} & 3.4 & 0.80 $\pm$ 0.10  \\
NUSEX\cite{ref:nusex} & 0.74  & $0.96^{+0.32}_{-0.28}$ \\
Kamiokande\cite{ref:kammu} & 7.7 & $0.60^{+0.06}_{-0.05} \pm 0.05$ sub-GeV \\
           &     & $0.57^{+0.08}_{-0.07}\pm 0.07$ multi-GeV \\
Frejus\cite{ref:frejus} & 1.56 & $0.95 \pm 0.18 $ \\
\hline     
\end{tabular}
\caption{\label{ta:atmother} Summary of measurements of flavor
           composition of atmospheric neutrinos from underground experiments.}
\end{center}
\end{table}

\subsubsection{Upward Going Muons}

Measurement of the flux of upward going muons also provides
information about atmospheric neutrino oscillations.  Neutrino
interactions in the rock surrounding the detector produce muons
energetic enough (energies above a few GeV) to pass through the
detector.  The direction of the scattered lepton roughly approximates
the neutrino direction. 
However, since only muons are detected, the
extraction of an oscillation signal relies 
on the calculation of the normalization and angular distribution of the
flux via Monte Carlo and is thus more prone to systematic uncertainty.

SuperKamiokande \cite{ref:skmuons} has measured a total of 614 upward through going
muons with energies above 1.6 GeV. The measured flux agrees with
calculation within the (large) errors. The zenith angle distribution
peaks more sharply at the horizon ($\cos \theta_{Z} \sim 0$) than
expected for the no-oscillation case.  The best fit gives 
$10^{-3} < \delta M^2 < 10^{-1}$ eV$^2$ for $\sin^2 2 \theta > 0.4$.  The flux is
14\% lower than predicted in \cite{ref:atnucalc}.

A measurement of the upward going muon flux by the MACRO experiment \cite{ref:macro}
also differs from the no oscillation expectation.
MACRO, a 5 kiloton detector, uses streamer tubes and scintillator to
measure the trajectory and velocity with 1 cm and 1 ns precision,
respectively, allowing the measurement of upward through going muons
produced by neutrinos with energies $E_{\nu}=1-100$ GeV. A total of 451 
$\nu$ induced events are measured (479 with a background of $28 \pm
7$) while 612 are expected and, as with SuperKamiokande, the zenith angle distribution
peaks somewhat more at the horizon.  A fit to the zenith angle distribution gives $2
\times 10^{-4} < \delta M^2 < 2 \times 10^{-2}$ eV$^2$ for $\sin^2 2 \theta > 0.5$.  


\subsubsection{Summary}

In summary, a deficit of $\numu$ induced events appears in almost all of the
measurements, the most precise of which comes from SuperKamiokande.  In 
addition, SuperKamiokande measures a zenith angle dependent $\numu$ flux
which, although statistically less significant, is less prone to theoretical
uncertainty than the flux measurement. The data are well fit by the hypothesis of
neutrino oscillation. The measurement of the upward
going muon flux and the agreement between the predicted and measured azimuthal
$\numu$ induced event asymmetry provide important supporting evidence in
favor of this hypothesis.


\subsection{Solar Neutrinos}\label{sect:solar}\label{sec42}

\begin{figure}
\hspace*{0.1\textwidth}\epsfxsize=0.8\textwidth\epsfbox{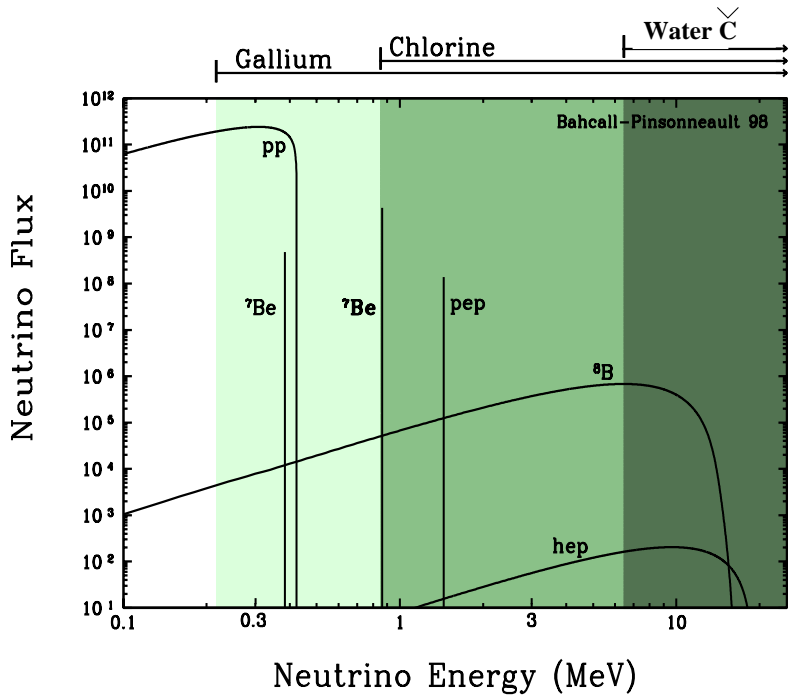}
\caption{\label{fi:solarnu} Energy spectrum at earth of solar neutrinos.  The
flux for lines is given as neutrinos cm$^{-2}$-s.  The continuous spectra
are in units of neutrino cm$^{-2}$/MeV/s.  From \cite{ref:bpweb}. }
\end{figure}

\newcommand{\boron}{^8\mathrm{B}}
The sun produces electron neutrinos via nine principal reactions in the $pp$
chain, starting with the fusion of two protons to make a deuteron and
ending with the production of $^4\mathrm{He}$ \cite{ref:bahcallbook}. 
The neutrino energies range
from 0 to 15 MeV and three of the fusion reactions result in
mono-energetic $\nue$'s at 380, 860 and 1400 keV,
Figure~\ref{fi:solarnu}.  
The calculation of the solar neutrino flux \cite{ref:bstand} 
has an uncertainty of less than 3\%, and therefore one can probe for
neutrino oscillation by searching for $\nue$ disappearance.

\subsubsection{Experiments and Results}

Three different types of experiments have measured the solar neutrino
flux.  Owing to the different experimental techniques, each measures a
different portion of the neutrino energy spectrum.  Water \v{C}erenkov
\cite{ref:imb, ref:kam2, ref:sksolar, ref:skspectrum}
counters detect neutrinos via $\nue + e^- \ra \nue + e^-$ and, owing
to their relatively high energy threshold of 5-7 MeV,
detect only the upper part of the $\boron$ $\nue$
spectrum.  The original solar neutrino experiment \cite{ref:davis} 
relies on the reaction $\nue + ^{37}$Cl $ \ra ^{37}$Ar$ + e^-$ in perchloroethylene,
which has a threshold of 814 keV,
giving sensitivity to the $^8$B, $^{7}$Be, and $pep$ neutrinos and the upper part of
the CNO spectrum.  Finally, the SAGE \cite{ref:sage} and Gallex
\cite{ref:gallex} experiments use the capture
reaction $\nue + ^{71}$Ga $\ra ^{71}$Ge$+ e^-$ in metallic gallium
(SAGE) or gallium chloride (Gallex) with a threshold of 232 keV.  This
provides a window for detection of the $pp$ $\nue$s.
The fluxes observed by the various experiments are compared in
Table~\ref{ta:solarresults} with the values predicted assuming that neutrinos do not
oscillate.

These experiments have very different systematic effects. The
\mbox{$\nue + e^- \ra \nue + e^-$} elastic scattering used by the water
\v{C}erenkov experiments gives a correlation (within about 25$^{\mathrm{o}}$) between
the recoil electron and the incident neutrino direction, allowing the
rejection of the isotropic backgrounds from the gamma ray photons emitted by
radio-contaminants in the detector itself.  Decay products from $^{16}$O
created by spallation reactions induced by cosmic ray muons are also a
serious background and are rejected via time and spatial correlation
with the track from the cosmic ray muon entering the detector.  
The clear correlation between the sun's location and the incident neutrinos
(see \cite{ref:sksolar}, Figure~2) gives a large measure of comfort that the observed
events are the result of stellar processes.

The other three experiments rely on the exposure of a large target mass to
the $\nue$ from the sun, which convert a few atoms per day in the target
to a radioactive daughter (via induced beta decay), 
which is then removed from the target mass
chemically and detected via the x-ray from electron capture induced decay.  
For both the chlorine and gallium experiments, 
the daughter nucleus was extracted as a gas
and detected in proportional counters via the subsequent electron
capture decay.  The collection efficiency is
close to 100\% and detection efficiency around 50\%.  Background enters 
primarily through the counting procedure where cosmic ray muons
and gammas from radio-isotope contamination (primarily from airborne radon)
contribute about 10\% to the overall counting rate.  While the nature
of the experiments does not allow direct verification that the neutrinos
actually come from the sun, both gallium experiments performed
calibrations using a $^{51}$Cr \cite{ref:gacalib} source which provided
751 keV $\nue$ via
$e^- + ^{51}$Cr $\ra ^{51}$V $ + \nue$.  In both cases the calibration
confirmed the expected collection and counting efficiencies, with 
$ R_{calib/calc}=83 \pm 10\%$ for Gallex and $95 \pm 11^{+5}_{-8}$ for
SAGE.

\begin{table}
\begin{center}
\begin{tabular}{|c c c c c|}
\hline
Target   & Experiments & Observed/Expected Rate & Source \\
\hline
$^{37}$Cl & Homestake & 0.331 $^{+0.061}_{-0.053}$  & $^8$B,$^7$Be   \\
$^{71}$Ga & SAGE      & 0.519 $^{+0.070}_{-0.066}$  & $pp, pep,^7$Be \\
          & Gallex    & 0.605 $^{+0.060}_{-0.054}$  &                \\
H$_2$0    & SuperKamiokande & 0.470 $^{+0.061}_{-0.054}$ & $^8$B \\      
          & Kamiokande & 0.56 $\;^{+0.091}_{-0.054}$       & \\
\hline
\end{tabular}
\caption{\label{ta:solarresults} Results from solar neutrino
experiments.  Expected fluxes are from \cite{ref:bp98}.}
\end{center}
\end{table}

The original observation of solar neutrinos was made by the Homestake
experiment \cite{ref:davis}, which has been in operation almost
continuously since 1970. The target consists of 615 metric tons of
C$_2$Cl$_4$, which provides about $2 \times 10^{30}$ target $^{37}$Cl
atoms.  Solar neutrinos with energies above 814 keV are absorbed via
the reaction $\nue + ^{37}$Cl $\ra ^{37}$Ar $+ e^-$ to give
$^{37}$Ar which decay via electron capture with a half-life of 35 days.
The measurement is performed as a series of exposures of four to six
weeks during which time the concentration of $^{37}$Ar builds up to
saturation.  The $^{37}$Ar is then extracted by bubbling helium
through the tank and chemically separating the $^{37}$Ar into a
proportional tube where the number of $^{37}$Ar atoms are counted via
the detection of the Auger electron from their electron capture
decay.  

In addition to a measurement of the total flux of solar neutrinos
above 5.5 MeV, SuperKamiokande \cite{ref:skspectrum} 
has measured the recoil energy spectrum
of electrons above 5.5 MeV, which gives an indication of the energy
spectrum of solar neutrinos.  The energy scale is calibrated to a
precision of 0.8\% at 10 MeV using a combination of a dedicated LINAC,
$^{16}$N (from muon spallation) and muon decay.  The resulting
spectrum is harder than expected even for the MSW oscillation result
favored by the total rate measurements.  
An earlier, much less precise measurement \cite{ref:kam2}
showed no disagreement with the expected flux.

SuperKamiokande has also measured the variation of the solar neutrino
flux depending on whether the neutrinos from the sun pass through the
earth or not (the ``day/night'' effect)\cite{ref:skdaynight}.
Day/night variation could be caused by MSW oscillations in the earth.
The measured flux ratio is 
\begin{equation}
\frac{\Phi_{night}}{\Phi_{day}} = 1 +0.047 \pm 0.042 \mathrm{( stat.)}
\pm 0.008 \mathrm{( syst.)}
\end{equation}
indicating no day/night effect at the 10\% level.

\subsubsection{Combined Analysis and Future Probes} 

Several combined analyses of all the solar neutrino data 
have been carried out \cite{ref:bstand,ref:bp98}.  Perhaps the most impressive allows
normalizations of the
$pp$, $^{7}$Be, $^8$B and CNO neutrino spectra to vary while assuming that 
neutrinos do not oscillate.  No combination of normalizations
reproduces the observations with significant compatibility, indicating that 
some new phenomenon, consistent with neutrino oscillations, takes
place regardless of the details of the solar model. If one assumes that solar
neutrinos do undergo oscillations, either of the MSW variety within the sun,
or of the vacuum variety between the sun and the earth, then a good
fit to all the solar neutrino data, apart from the new data on the energy spectrum,
can be obtained. The best fit global solutions are summarized in
Table~\ref{ta:mswsolution}. Inclusion of the limit on the day/night variation tends to
push the lower limit on $\delta M^2$ to above 10$^{-5}$ eV$^2$ for the large-angle MSW
solutions \cite{ref:skdaynight,b29}. 

\begin{table}
\begin{center}
\begin{tabular}{| l l l |}
\hline
Type      & Large angle   & Small angle \\
\hline
MSW $\nue\ra\nu_{\mathrm Active}$ &  $\delta M^2 = 1.8 \times 10^{-6} \mathrm{eV}^2 $
&
$\delta M^2 = 5.4 \times 10^{-6} \mathrm{eV}^2$ \\
                  &   $\sin^2 2 \theta = 0.76$ &
$\sin^2 2 \theta = 6.0 \times 10^{-3}$ \\
\hline
MSW $\nue\ra\nu_{\mathrm Sterile}$ &  &$\delta M^2 = 4.3 \times 10^{-6} \mathrm{eV}^2
$ \\
                  &  & $\sin^2 2 \theta = 6.9 \times 10^{-3}$  \\
\hline
Vacuum           & $\delta M^2=8.0 \times 10^{-11} \mathrm{eV}^2$ & \\
                  & $\sin^2 2 \theta=0.75 $ &  \\
\hline
\end{tabular}
\caption{\label{ta:mswsolution} Best fit results to all solar
                  neutrino flux data \cite{ref:bstand}. 
}
\end{center}
\end{table}

All the results in Table~\ref{ta:mswsolution} are in some contradiction
with the SuperKamiokande recoil spectrum measurement, with the largest
deviation being at the highest electron recoil energies. It will be interesting to see
how this situation evolves as the spectrum measurements become more precise.

Whether the electron neutrinos produced in the sun oscillate
to active ($\numu$ or $\nu_\tau$) or sterile neutrinos will be largely resolved by the
Sudbury Neutrino Observatory (SNO) \cite{ref:sno} experiment, which will begin
operation next year with a target of 1 kiloton of heavy water
(D$_{2}$O). The use of a heavy water target allows the simultaneous 
measurement of the charged current and neutral current cross sections via
\begin{eqnarray}
    \nue + d & \ra & p+p+e^- \mathrm{  (CC)} \\
    \nu_{x} + d & \ra & p+n+\nu_{x} \mathrm{  (NC) ,} 
\end{eqnarray} 
where $\nu_{x}$ refers to $\nue$, $\numu$ or $\nu_\tau$.
The first reaction will measure the flux of solar $\nue$, while the second reaction
will measure the total flux of $\nue, \numu$
and $\nu_{\tau}$. If the MSW effect or vacuum oscillation converts
solar electron neutrinos into sterile neutrinos, the rate for the
second reaction will be lower than if the conversion is to active
neutrinos. The SNO experiment will be sensitive only to $^8$B neutrinos. The 
rates expected assuming 50\% of the solar neutrinos oscillate are
5.5 NC events per day (assuming 40\% detection efficiency) and 12.7
CC events per day. 
Thus, the experiment should resolve the situation after two years of
operation. The heavy water target is surrounded by a 7800 ton pure
water buffer and is viewed by 9456 photo-multiplier tubes.  The
\v{C}erenkov ring from each final state charged particle gives its direction, and the
number of \v{C}erenkov photons gives the particle's
total energy.  The neutrons produced in the neutral current scattering
are detected by delayed coincidence; they first thermalize in the
target and are subsequently captured on a proton, giving a 2.2 MeV
gamma ray.

In the coming years, SNO, SuperKamiokande and BOREXINO will make
important inroads in the measurement of the solar neutrino energy
spectrum.  In particular, BOREXINO \cite{ref:borexino}, 
located in the Gran Sasso, will
measure the $^7$Be neutrino flux using $\nue + e^- \ra \nue + e^-$.
The detector consists of 300 tons of very high radio-purity scintillator
(less than 10$^{-14}$ g of U, Th and Rn per gram of scintillator) viewed by 2000
photo-multiplier tubes.  The arrival time of the scintillation photons
at the PMT will be used to locate the event vertex inside a 100 ton
fiducial volume and the total signal from all PMTs will give the
energy of the recoil electron.  BOREXINO plans to achieve a 250 keV
threshold, which will allow the detection of $^7$Be $\nue$ which will
create recoil electrons up to 660 keV.  For the standard solar model
with no oscillations, the count rate is expected to be 55 events per day.
If the MSW effect is operative, the count rate would be reduced, and its
value may discriminate between the large and small angle MSW solutions.
In addition, the high statistics of BOREXINO will allow the
seasonal variation to be measured, which is expected to be large if
vacuum oscillations are taking place.


The next few years should see a resolution of the solar neutrino
problem: we can expect measurements of the $\nue$ energy spectrum at
several points which will distinguish between the small- and large-angle MSW
possibilities. Measurements of the seasonal variation will determine if
vacuum oscillations are taking place.  Finally, comparison of the
neutral and charged current rates will clarify the role of sterile
neutrinos.


\subsection{The LSND Experiment}\label{sect:LSND}

To date there are only two claimed positive observations of neutrino
oscillations from an accelerator-based neutrino source, both from the LSND
(Liquid Scintillator Neutrino Detector) experiment at the LAMPF neutrino
source.  LSND has not only the sole observation of oscillations
from human-made neutrinos, but also the only appearance signal among our
three hints of oscillations.  The LAMPF accelerator produces a large current
of nearly relativistic protons, and a large number of $\pi^+$ are produced in
a long water target upstream of a copper beam stop.  Most of these protons
come to rest and then decay {\em via} the familiar chain,
$\pi^+\to\mu^+\nu_\mu$, $\mu^+\to e^+\nu_e\nub_\mu$, where the muon decays at
rest.  The charge conjugate chain beginning from $\pi^-$ is suppressed by
approximately three orders of magnitude, roughly one each from production,
absorption of $\pi^-$ in the beam stop, and atomic capture of $\mu^-$ before
decay \cite{LSND DAR}.  Therefore, a $\nub_e$ appearance signal is relatively
free of beam-induced backgrounds.  LSND also searches for the appearance of
high-energy $\nu_e$ (above the endpoint of $\mu^+\to e^+\nub_\mu\nu_e$ decay
at rest), which could result from $\pi^+\to\nu_\mu\mu^+$ in flight, and
$\nu_\mu\to\nu_e$ \cite{LSND DIF}.  The experiment observes significant
excesses of both $\nub_e$ and $\nu_e$, which it attributes to
$\nub_\mu\to\nub_e$ from muon decay at rest and $\nu_\mu\to\nu_e$
oscillations from pion decay in flight, respectively.  The $\nub_e$
appearance signal is more significant than the $\nu_e$ signal because of
higher statistics, and lower beam-related and detector-related backgrounds.
 
\begin{figure}
\epsfxsize=\textwidth\epsfbox{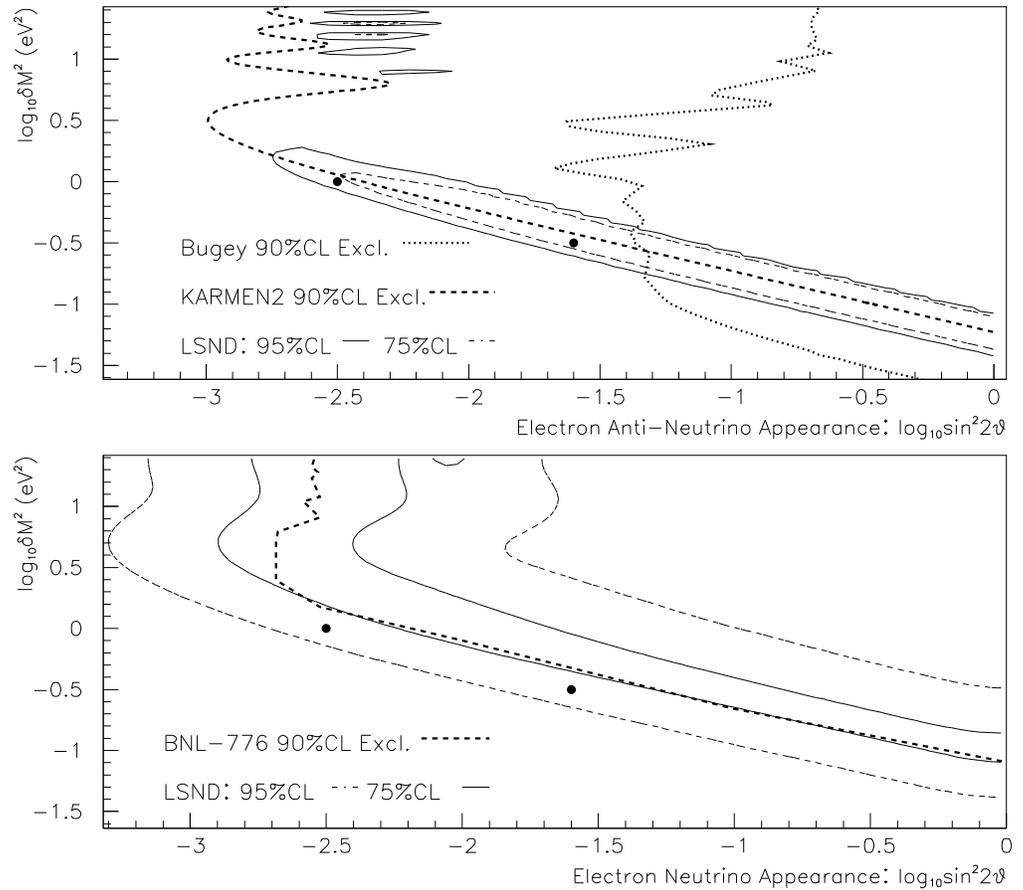}
\caption{The $95$\% C.L.\ regions for $\nub_\mu\to\nub_e$ and
  $\nu_\mu\to\nu_e$ oscillations from the LSND $\nub_e$ and $\nu_e$ excesses.
  {\em (Data provided courtesy of the LSND collaboration.)}  Shown also are
  the 90\% C.L.\ exclusion curves for the ongoing KARMEN2 experiment
  {\em (preliminary limit provided by
  courtesy of R.~Maschuw)} as of January 1999, for the Bugey reactor
  experiment, and for BNL-E776.  Higher $\delta M^2$ are ruled out by CCFR,
  NOMAD and NuTeV data.  Two points in the LSND-favored region from the
  $\nuebar$ appearance analysis are shown to guide the eye.}
\label{fig:lsnd-dar-dif}
\end{figure}

The two favored regions in the space of two generation oscillation parameters
for the interpretation of these excesses are shown in
Figure~\ref{fig:lsnd-dar-dif}.  The two favored regions in this space are
consistent.  After background subtraction, the mean energy of the 
excess $\nub_e$ is roughly a factor of two lower than the mean energy of the
excess $\nu_e$, and thus the two
appearance experiments are sensitive to different $\dmsq$.  The high $\dmsq$
oscillation probabilities from $\nub_\mu(\nu_\mu)\to\nub_e(\nu_e)$ in the two
analyses are presented as $(3.1^{+1.1}_{-1.0}\pm0.5)\times10^{-3}$ and
$(2.6\pm1.0\pm0.5)\times10^{-3}$, respectively.  The $\nu_e$ analysis
observes an excess with a somewhat harder energy spectrum than $\nu_\mu$
oscillations might predict and therefore weakly favors high $\delta M^2$.
The $\nuebar$ analysis strongly favors low $\delta M^2$.

On the face of it, LSND's statistical evidence, at least for the $\nuebar$
appearance observation, is overwhelming.  Suggestions have been made that one
portion of the $\nuebar$ suppression in the $\pi^-$ decay-at-rest chain could
be significantly altered, by, for example, meta-stable $\pi^-$-$N$ bound
states, but there appears to be no compelling case for proving such a
hypothesis.  Another alternative to an oscillation explanation is the
production of unexpected $\nuebar$ in the LSND decay chain {\em via}
lepton-flavor violating processes in left-right symmetric models, in R-parity
violating SUSY and in model-independent frameworks \cite{LFV-LSND}.  Such
processes are strongly constrained by existing data on rare muon decays and
on muonium-antimuonium conversion, and do not appear to yield a convincing
non-oscillation explanation at this time.

\subsubsection{LSND AND COMPETING NULL RESULTS}

While LSND is the only accelerator experiment to observe oscillations to
date, many other experiments have probed oscillations in regions of the
favored LSND oscillation parameter space.  These competing experiments can
be grouped into three categories: experiments with a similar beam and energy
range (KARMEN and KARMEN2), reactor experiments and high-energy experiments.

The KARMEN and KARMEN2 experiments at the ISIS neutron spallation facility
use a beam, like the LSND beam, produced primarily from pion decay and muon
decay at rest.  However, the $100$~ns pulsed KARMEN beam has a very low duty
cycle ($\sim10^{-5}$) and allows for time-based separation of $\pi^+$ and
$\mu^+$ decay products.  The KARMEN experiment had a significant background
from neutrons produced in the shielding which limited its sensitivity;
however, the KARMEN2 upgrade has reduced this background enormously
\cite{KARMEN-old}.

Based on a nearly $3000$~Coulomb exposure, KARMEN2 initially reported a
signal of $0$ $\nu_e$ candidate events where a background of $2.88\pm0.13$
events was expected \cite{KARMEN-old}.  However, an updated analysis based 
on nearly
$4000$~Coulombs with looser cuts now reports $7$ observed events with an
expected background of $9.3$ events \cite{KARMEN-update}.  Despite seeing no
signal, KARMEN2 currently lacks the sensitivity to convincingly rule out the
LSND observation.  However, should KARMEN2 improve its sensitivity and
continue to see no evidence of oscillation into $\nuebar$, it may call into 
question the reproducibility of the LSND $\nuebar$ excess, since the beams and
detection mechanisms are so similar.

Reactor experiments, which are discussed in more detail in
Section~\ref{sect-react}, are disappearance experiments at low neutrino
energy and are therefore sensitive to high $\sin^22\theta$ and low $\delta
M^2$.  The observation of no oscillation signal from the Bugey experiment in
the $\nuebar\to\nuebar$ channel limits the possible $\numubar\to\nuebar$
probability, and therefore excludes the LSND favored region at high mixing
angle.

High-energy accelerator experiments currently suffer from generally
insufficient $L/E$ to probe the LSND $\delta M^2$ ``knee'' in
Figure~\ref{fig:lsnd-dar-dif}.  Most accelerator
experiments also probe primarily $\numu\to\nue$, although the NuTeV
experiment expects to be sensitive to the LSND region above $\delta M^2$ of
$\sim 20$~eV$^2$ in its $\numubar$ beam \cite{Harris-nuebar-osc}.  CCFR
\cite{CCFR-Romosan} and NOMAD \cite{NOMAD-nue} rule out the LSND favored
region above $25$ and $10$~eV$^2$, respectively, and the result from BNL-E734
\cite{BNL-E734} rules out much of the LSND favored oscillation parameters in
the same $\delta M^2$ region.  BNL-E776, with its sensitivity to larger $L/E$
comes the closest of any existing experiment to overlapping the LSND favored
region \cite{BNL-E776}.  Like KARMEN2, BNL-E776 cannot probe the entire LSND
favored region in the $\nuebar$ signal at lowest $\delta M^2$.  However, the
BNL-E776 result does exclude the most probable values from the LSND
decay-in-flight analysis, as shown in Figure~\ref{fig:lsnd-dar-dif}.

\subsubsection{FUTURE TESTS OF THE LSND RESULT}

Future accelerator experiments are planned to search for oscillations in the
LSND favored region.  An experiment, BooNE \cite{BooNE}, has been approved
for single-detector operation at Fermilab, and an experiment at CERN with a
similar baseline and beam energy, I-216 \cite{I-216}, is also being proposed.
Both experiments hope ultimately to use neutrino beams of about $1$~GeV to
study $\nu_e$ appearance in detectors at two different values of $L$.
Because of the energy regime, both experiments expect large detector and beam
backgrounds to $\nu_e$ appearance; however, this next generation of
experiments should have the raw statistical power to convincingly address the
entire LSND favored region.


\section{A COSMOLOGICAL CONSTRAINT ON NEUTRINO MASS}\label{sect:lambda}

We now turn from the positive hints of neutrino mass to a discussion of the
constraints on this mass from cosmology and from negative laboratory searches.

Owing to the great abundance of neutrinos in the universe, even tiny neutrino
masses can lead to a significant neutrino contribution to the mass density of
the universe. As a result, our information on this mass density provides a
fairly stringent upper bound on neutrino mass.

Increasing evidence from several sources suggests that \cite{br5,br6} 
\beq
\Omega_M = 0.1 - 0.4 \;, \label{eqc1} 
\eeq 
where $\Omega_M$ is the total mass
density of the universe, as a fraction of the critical density required to
stop indefinite expansion. This $\Omega_M$ includes a contribution from
baryons. Very likely, it also includes a contribution from non-baryonic Cold
Dark Matter (CDM) particles, which are much heavier than neutrinos. These
heavier particles appear to be needed to help seed the early-universe
formation of the observed galactic structure.  Finally, $\Omega_M$ may
include a contribution from Hot Dark Matter (HDM) particles. These are
long-lived particles with masses much less than 1 keV, which would have been
relativistic at the early time important for galactic structure formation. If
neutrinos are massive, their contribution to $\Omega_M$, $\Omega_\nu$, is a
part of the HDM contribution.

The HDM part of the mass density cannot be too large, because in the early
universe any HDM would have acted as a drag interfering with the formation of
the observed galactic structure. From this fact, the bound $\Omega_M \; \lsim
\; 0.4$ (\Eq{c1}), and the fact that some of $\Omega_M$ comes from baryons
and from non-baryonic CDM, one may conclude conservatively \cite{br6} that
the HDM part of $\Omega_M$ does not exceed 0.2. This means that, in
particular, $\Omega_\nu \; \lsim \; 0.2$.

For any neutrino mass eigenstate $\nu_m$ which is a constituent of active
flavor neutrinos, the present number density in the universe is fixed at
115/cc by thermal equilibrium in an early era \cite{br7}. Thus, if $\nu_m$
has mass $M_m$, its present mass density is 115 $M_m$/cc. Now, the present
critical density is $1.05 \times 10^4 h^2$ eV/cc, where $h$ is the Hubble
constant in units of 100 km/sec/Mpc. Hence, assuming $h \simeq 0.65$, the
contribution of $\nu_m$ to $\Omega_\nu$ is $M_m$/40 eV. If $\nu_e$,
$\nu_\mu$, and $\nu_\tau$ are made up out of three mass eigenstates with
masses $M_1$, $M_2$, and $M_3$, these mass eigenstates together yield 
\beq
\Omega_\nu = (M_1 + M_2 + M_3) / 40 \mbox{ eV .}
\label{eqc2}
\eeq 
Requiring that $\Omega_\nu \;\lsim \;0.2$, we obtain the cosmological upper
bound on neutrino mass 
\beq M_1 + M_2 + M_3 \; \lsim \; 8 \mbox{ eV .}
\label{eqc3}\label{eqHDM}
\eeq

\section{NEGATIVE SEARCHES FOR NEUTRINO MASS}

\subsection{Reactor Oscillation Searches}\label{sect:react}\label{sect-react}

Nuclear power reactors have always played a central role in neutrino
physics.  Neutrinos were first observed at a reactor \cite{ref:reines} using
\begin{equation}
\nuebar + p \ra n + e^+.
\end{equation}
The positron is detected via its ionization energy followed by
detection of the Compton scattering of the two 511 keV gammas from the
annihilation with an electron at rest.  The neutron thermalizes and is
then captured by a stable isotope (in this case cadmium) with a high
capture cross section to form a gamma emitter whose gammas are then
detected.  This gives a double delayed coincidence which provides a
powerful means for background rejection.  Reactors emit $8 \times
10^{20} \nu$/s per gigawatt (electrical) power, so a one ton target
mass 100 meters from a reactor core will detect a few neutrino events
per hour.  The neutrinos emitted by the reactor have an average energy
of 3 MeV, so this type of experiment allows the probing of the $\delta M^2 \sim
10^{-2}$ eV$^2$ region.  Owing to the low neutrino
energy, only $\nuebar$ disappearance experiments are possible, restricting
the sensitivity to the mixing angle to $\sin^2 2\theta>0.1$.

Two first generation reactor neutrino oscillation experiments probed
the $\delta M^2 \sim 10^{-2}$eV$^2$ range \cite{ref:gandb}.  
Both were one ton target
masses at $L=100$m and used $^3$He filled wire chambers to detect the
neutron, giving efficiencies of around 15-20\%.  In both experiments, data
was taken at different distances from the reactor, allowing two
different measurements.  The first was the comparison of the positron
energy spectrum between different distances;  the oscillation
pattern depends on $L/E$, so changing $L$ would change the neutrino
energy spectrum and thence the positron energy spectrum.  This allowed
the measurement to be performed with no {\em a priori} knowledge of
the neutrino energy spectrum from the reactor.  In the second
analysis, the knowledge of the fuel composition in the reactor core
was used to calculate the expected flux which could then be used to
compute the expected positron energy spectrum at a given distance.
Different calculations of the flux were found to agree at the 3\%
level with dedicated measurements \cite{ref:reactflux}, 
removing the necessity of moving the detectors in future,
larger experiments.

The current generation of reactor oscillation experiments, Chooz \cite{ref:chooz} and
Palo Verde \cite{ref:pv}, have target masses of several tons (5 t and 12 t, 
respectively) and are located roughly 1 km (0.74 km and 1 km,
respectively) from the several reactors.  Both use liquid scintillator
targets loaded with 0.1\% Gd (by weight), which has a high neutron
capture cross section and which radiates 8 MeV of gamma rays after
capture.  This detection scheme gives a significantly higher detection
efficiency for the neutrons, resulting in higher performance overall.

\begin{figure}
\epsfxsize=0.6\textwidth\hspace*{0.2\textwidth}\epsfbox{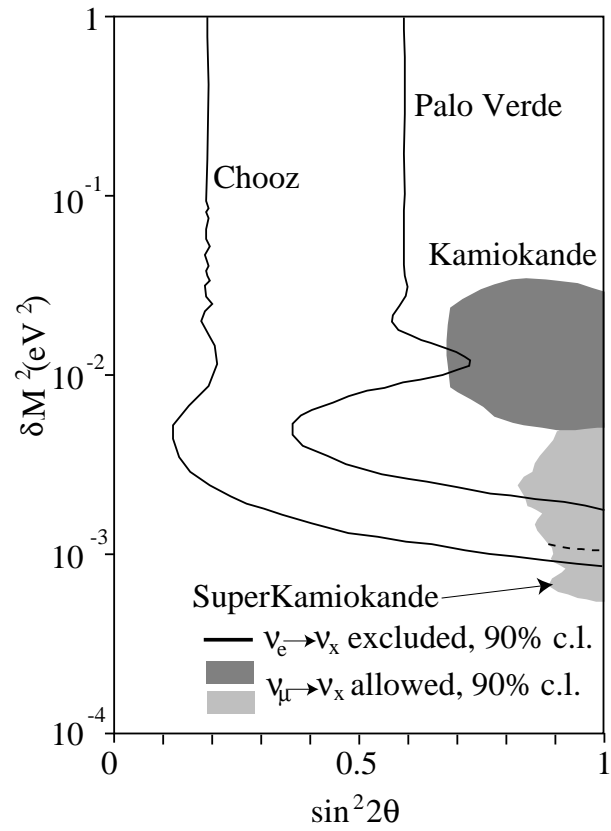}
\caption{\label{fi:excl} Exclusion plot for reactor neutrino
experiments. Also shown is the allowed region from the atmospheric
experiments.  The dotted line in the SuperKamiokande region indicates a lower
bound on $\dmsq$ obtained from Ref.~\cite{ref:skmuons}, an analysis of
upward-going muon data.}
\end{figure}

Both experiments report null results and rule out $\nuebar \ra \overline{\nu}_x$
oscillations for $\delta M^2 > 9 \times 10^{-4}$eV$^2$ and $\sin^2 2
\theta > $ 0.1,  Figure~\ref{fi:excl}. 
Combined with the atmospheric
neutrino measurements from SuperKamkiokande, the reactor measurements
indicate that at the 90\% confidence level, the atmospheric $\numu$ oscillate into
$\nutau$ or a sterile neutrino, rather than into $\nue$.
Both reactor experiments are continuing to run and will eventually reach 
$\delta M^2 \sim 10^{-4}$eV$^2$ for large mixing angles. If no oscillation
effect emerges, they will completely exclude $\numu \ra \nue$ as the primary
oscillation process observed by SuperKamiokande.

A very ambitious experiment, KAMLAND \cite{ref:kamland}, seeks to use
the old Kamiokande site to situate a 1 kiloton detector 150 km from a complex of
several reactors, giving a sensitivity of $\delta M^2 \sim 2 \times 10^{-5}$
eV$^2$. As KAMLAND is a disappearance experiment, it would only be
sensitive to $\sin^2 2\theta > 0.1$ and would effectively resolve
between the large and small angle MSW solutions to the solar neutrino measurements.
KAMLAND would use a large scintillator target viewed by roughly 2000
phototubes, giving a signal rate of two events per day. It is thought that through
careful selection of construction materials and purification of the
scintillator target, the background rate could be reduced to 0.1 events per day. 
KAMLAND is planning to take data for three years starting in 2001.

The scattering of reactor neutrinos from electrons at low energies may be
used to measure the neutrino magnetic moment $\mu_{\nu}$.  
In the standard model, $\mu_{\nu}\sim 10^{-19}
M_{\nu}(\mathrm{eV})\: \mu_B$ for Dirac neutrinos and $\mu_{\nu}=0$ for
Majorana neutrinos; in both cases too small to explain any of observed
oscillation effects.  The current laboratory limit is 
$\mu_{\nu} < 10^{-10}\mu_B$ at 95\%
c.l. \cite{ref:derbin}.  In the near future, the MUNU collaboration
plans to use a 1 m$^3$ TPC with a recoil electron threshold near 100
keV to push the sensitivity to $3 \times 10^{-11} \mu_B$ \cite{ref:munu}.


\subsection{Accelerator Oscillation Searches}\label{sect:accel}

Aside from the appearance signal at the LSND experiment that was discussed in
Section~\ref{sect:LSND}, there is no other evidence for neutrino oscillations
observed in accelerator-based experiments.  Nevertheless, accelerator
experiments are the preferred venue for studying neutrino oscillation
phenomena because of the ability to control and vary $L/E$ (and thus $\delta
M^2$ sensitivity) in a neutrino beam and the ability to precisely determine
flavor content of the neutrino beam at the point of creation.  However, the
practicalities of accelerator experiments place severe restrictions on our
ability to probe the $L/E$ regions favored by current non-accelerator
observations of oscillation.  In this Section, we summarize the current
capabilities and status of accelerator experiments, as well as future
expectations.

High intensity neutrino beams at accelerators are typically produced by high
energy hadronic interactions in a fixed target which copiously produce
forward secondary mesons, $\pi^\pm$ and $K^\pm$ in particular.  These mesons
can then be captured and focused parallel in electromagnetic beam optics,
where the mesons are then allowed to decay weakly to a tertiary ``beam'' of
neutrinos, predominantly $\numu$ and $\numubar$.  The efficiency for this
process varies with the type of beam optics and primary beam energy, but
$10^{-2}$--$10^{-5}$ neutrinos per incident hadron is a representative range.
Because the neutrinos cannot be subsequently focused, at long distances $L$
from the production point the neutrino flux varies with $L^{-2}$.
Experiments may be roughly divided into two categories, short-baseline
experiments which locate their detectors as close as possible to the point of
neutrino production to maximize statistics, and long-baseline experiments
which pick a longer $L$ in order to tune $L/E$ sensitivity to the desired
range.  Contemporary high energy neutrino beams optimized for statistics run
at beam energies from a few $GeV$ to a few hundred $GeV$, and shielding
concerns require an $L$ of $100$~m--$1$~km, thus setting the ``natural''
range of $\delta M^2$ sensitivity at $10$--$100$~eV$^2$.  Lower $\delta M^2$
sensitivity, made interesting by the solar and atmospheric neutrino
oscillation signals, requires longer $L$ and/or lower $E$.  Lower $E$ allows
the use of more intense primary beam sources to generate the beam, but
decreases the neutrino cross-section (roughly proportional to $\sqrt{s}$) and
increases the beam divergence (roughly proportional to $\gamma^{-1}$, the
inverse Lorentz factor for the weakly-decaying parent particle).
Charged-current production of $\tau$ leptons from nuclear targets has a
neutrino energy threshold of $m_\tau+m_\tau^2/2m_N\sim3.3$~GeV and suffers
from significant mass suppression until $E$ is far above this threshold.
Longer $L$ results in an $L^{-2}$ decrease in statistics as noted above.  The
argument has been made \cite{MGoodman?} that in a long-baseline appearance
experiment the number of observed wrong-flavor neutrinos when $\delta
M^2$(eV$^2)L/E$(km/GeV)$\ll 1$ is constant with $L$; however, it should be
noted that backgrounds will not be constant with $L$ and that an experiment
with $\delta M^2 L/E \ll 1$ leaves no possibility
to measure $\delta M^2$.  Some
neutrino beam and detector parameters for current and planned experiments are
summarized in Table~\ref{tab:accel-param}.

\begin{table}
\begin{tabular}{|l|l|l|l|l|}\hline
  Experiment & $\left< E_\nu \right> $ & Baseline & Fiducial Mass & 
  $\numu(\overline{\numu})$~CC/yr \\\hline
  CCFR/NuTeV & $100$~GeV & $1$~km &  $350$~ton & $10^6$ \\
  CHORUS & $30$~GeV & $0.6$~km &  $0.8$~ton & $3\times10^4$ \\
  NOMAD & $30$~GeV & $0.6$~km &  $3$~ton & $10^5$ \\
  MINOS & $5$--$15$~GeV  & $730$~km & $3.3$~kton & $10^4$--$3\times10^3$ \\ 
  K2K & $2$~GeV & $250$~km & $22$~kton & $5\times10^2$ \\  
  I-216 & $1.5$~GeV & $1$~km & $150$~ton & $5\times10^4$\\
 \hline
\end{tabular}
  \caption{Approximate neutrino beam and detector parameters for a selection
    of current and near future accelerator-based neutrino oscillation
    experiments.  The mean $E_\nu$ refers to observed neutrinos.}
  \label{tab:accel-param}
\end{table}

The most sensitive current results come from several generations of
experiments with relatively short baselines, and therefore little sensitivity
to the $\delta M^2$ of the three current neutrino oscillation observations.
The most sensitive $\nu_\mu\to\nu_\tau$ searches to date come from the CHORUS
and NOMAD experiments at CERN, which set upper limits in $\sin^2 2\theta$ of
$\sim 1$--$3\times10^{-3}$ at their peak sensitivity around $\delta M^2$ of $\sim
30$~eV$^2$, and which exclude $\dmsq$ above $\sim 2$~eV$^2$ at maximal
mixing \cite{CHORUS-nutau,NOMAD-nutau}.  Both experiments expect to improve
their mixing angle limits significantly when all their data is analyzed.
The most sensitive searches for $\nu_\mu\to\nu_e$ oscillation come from the
CCFR/NuTeV \cite{CCFR-Romosan}, NOMAD \cite{NOMAD-nue}, BNL-E734 \cite{BNL-E734}
and BNL-E776 \cite{BNL-E776} experiments, and these results were discussed in
Section~\ref{sect:LSND} and displayed in Figure~\ref{fig:lsnd-dar-dif}.

The next generation of oscillation experiments offers significant hope of
probing the oscillations reported by the LSND and
atmospheric neutrino experiments.  Experiments targeted on the former
(BooNE, I-216) were discussed in Section~\ref{sect:LSND}.  The long-baseline
experiments searching for evidence of the latter include MINOS at
Fermilab-Soudan \cite{MINOS}, K2K at KEK-Kamiokande \cite{K2K}, and
ICARUS \cite{ICARUS}, NOE \cite{NOE} and OPERA \cite{OPERA} at CERN-Gran Sasso.
By virtue of their $L/E$ ratio, each of these experiments has a peak
sensitivity at $\delta M^2\sim 10^{-2}$~eV$^2$.  Their $\nu_\tau$ appearance
capabilities differ substantially, however.  Most notably, by virtue of its
low energy beam, K2K only has the possibility to detect $\nu_\tau$ through
neutral current interactions, such as $\nu_\tau N\to\nu_\tau\pi^0X$.

Studies of CP violation in the atmospheric region of $\delta M^2$ are
potentially interesting because an observed CP asymmetry would indicate that
there are at least three neutrino mass eigenstates participating in the observed
oscillation.  Unfortunately, the most easily accessed signal for CP
violation, $P(\numu\to\nu_{e,\tau})\not=P(\numubar\to\nub_{e,\tau})$, can
be contaminated by the presence of ``fake CP'' asymmetries from the
interactions of the beam with the matter of the earth \cite{Bernabeu}.  A
cleaner signature would be evidence of T violation from
$P(\numu\to\nu_{e,\tau})\not=P(\nu_{e,\tau}\to\numu)$, which would not be
available to this generation of long baseline experiments.

In the far future, some of the cleanest and most intense neutrino
beams may come from a muon collider facility, either in the form of low
energy beams from the muon source, or from
$\mu^+\to e^+\nu_e\overline{\nu_\mu}$ decays.  An analysis of a beam
generated by decays in a $10$~GeV muon storage ring and observed on the other
side of the earth \cite{Geer} shows an $L/E$ reach $\sim10^3$~km/GeV with a
potential to probe low mixing probabilities even for modest muon production
rates.


\subsection{Kinematic Searches for Mass}

Unlike searches for neutrino oscillations, kinematic searches are sensitive
to non-zero neutrino masses in the absence of flavor or neutrino-antineutrino
mixing.  We discuss two important kinematic techniques which have placed
stringent bounds on the $\nuebar$ mass: $\beta$ decay endpoint experiments,
and the detection of neutrinos from Supernova 1987A.

\begin{figure}
\epsfbox{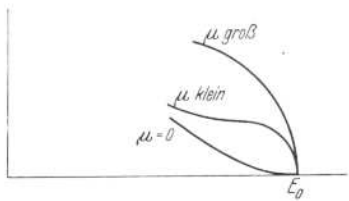}
\caption{\label{fi:fermi} Electron spectrum at endpoint
 \cite{ref:fermi} for differing neutrino masses $\mu$.}
\end{figure}

The original method for measuring the electron neutrino mass relies on the
precise measurement of the energy distribution of electrons emitted in
$\beta$ decay \cite{ref:fermi}. The number of counts as a function of
electron energy $E_e$ and momentum $p_e$ is \cite{ref:bandv}
\begin{equation}
\frac{dN}{dE_e} \propto F(Z,E_e) p_e E_e
(E_o-E_e)\sqrt{(E_o-E_e)^2-M_{\nu}^2} \; .
\end{equation}
Here, $F$ is the Fermi function, $Z$ the charge of the nucleus, and $E_o$ the Q value
of the decay neglecting the neutrino mass $M_\nu$. The neutrino mass is  
extracted from the  electron energy spectrum by fitting the
endpoint region of the spectrum with $M_{\nu}^2$ as a free parameter.
The square  of $M_\nu$ appears because the sensitivity to $M_{\nu}$ resides in the
curvature of the spectrum at the endpoint rather than the zero
crossing, Figure~\ref{fi:fermi}. The most sensitive experiments use the decay
\begin{equation}
^{3}H \ra ^{3}He + e^- + \nuebar \;,
\end{equation}
which has an endpoint energy of $18.5901\pm0.0017$ keV \cite{ref:the3}. This
elegant method is hampered by low count rates in the endpoint region (if
$M_{\nu}$ = 5 eV, the fractional change in the total number of counts is $ 3
\times 10^{-11}$) and detector resolution, which tends to smear counts to
higher energies, thus masking the effect of a massive neutrino.

Kinematic measurements require large sources and long running periods to accumulate
statistics near the endpoint, and spectrometer energy
resolutions better than 10 eV. In addition, atomic effects and electron
energy loss in the source and spectrometer must be understood and accounted for at the
few electron volt level \cite{ref:atomic}.  
Table~\ref{ta:kine} summarizes the measurements since 1990 and shows an asymptotic
approach to $M_{\nu} \sim 1 $ eV.

Somewhat disturbingly, the preponderance of the earlier experiments
shows a systematic trend of $M_{\nu}^2 < 0$, indicating either a
common systematic effect (for example in the calculation of the atomic
corrections) \cite{ref:atomexp} or a 
new physics effect \cite{ref:atomwacky}.  The most recent results from the
Troitsk and Mainz experiments somewhat mitigate this concern and give
limits of $M_{\nu} < 3$ eV.

Both the Mainz and Troitsk experiments propose ambitious upgrades to
probe the mass region below 1 eV in the next five years.  Such
measurements will begin to probe the mass region favored by the LSND
result and, in concert with further searches for double beta decay, help resolve the
properties of the electron neutrino. As illustrated in Figure~\ref{f8},
Section~\ref{sect:scena}, the neutrino mass eigenstate coupled most strongly to the
electron ($\nu_1$ in Figure~\ref{f8}) may have a mass too small to be seen in $^3$H
$\beta$ decay experiments. These experiments must then have negative results.

\begin{table}
\begin{center}
\begin{tabular}{| c c c c |}
\hline
Experiment   & $\Delta$E & $M_{\nu}^2$ & $M_{\nu}<$(95\% c.l.) \\
             &  (eV)     &  (eV$^2$)   &  (eV)  \\
\hline
LANL \cite{ref:lanl}      &  23   & $-148 \pm 68 \pm 41$  & 9.3      \\
Zurich \cite{ref:zurich}  &  17   & $-24  \pm 48 \pm 61$  & 11       \\
Tokyo INS \cite{ref:ins}  &  16   & $-65 \pm 85 \pm 61$   & 13       \\
\hline
Livermore \cite{ref:livermore} &           &                       & 8        \\
160 eV range &  18   & $-72 \pm 41 \pm 30$   &          \\
900 eV range &  18   & $-130 \pm 20$         &          \\
\hline
Mainz \cite{ref:mainz}    &           &                       &          \\
137 eV range &   6     & $-39 \pm 35 \pm 14$   & 7.2      \\
500 eV range &   6     & $-117 \pm 18$         &          \\
137 eV range &   6     & $-22 \pm 17 \pm 17$   & 6        \\
\hline
Troitsk \cite{ref:troitsk}   &           &                       &          \\
220 eV range &  4.3    & $-22 \pm 4.8$         & 1.7      \\
270 eV range &  4.3    & $-2.7 \pm 10.1 \pm 4.9$ & 4.5    \\
170 eV range &  3.7    & $3.8 \pm 7.4 \pm 2.8$ & 4.4      \\
Combined 94-6&           & $1.5 \pm 5.9 \pm 3.6$ & 3.8      \\
Combined 94-7&           & $-2.1 \pm 3.7 \pm 2.3$ & 3        \\
\hline 
\end{tabular}
\caption{\label{ta:kine} Summary of most recent kinematic searches
using tritium decay.  $\Delta E$ is the experimental energy resolution at the
endpoint.  $M_{\nu}^2$ is the value extracted from the fit
for the endpoint data, and $M_{\nu}$ the limit from the 
extracted value given by the
authors.  For the Livermore, Mainz and Troitsk experiments, the
results arising from fits to different energy ranges around the 
endpoint are given.
}
\end{center}
\end{table}  

Another type of kinematic constraint can be obtained from the observation of a
neutrino burst from a nearby supernova. Water \v{C}erenkov detectors can detect the
$\nuebar$s in such a burst. Typical models predict a burst of
$>10$~MeV $\nuebar$s concentrated in a $\sim5$~second time window.  By
studying the observed spread of arrival times as a function of $\nuebar$
energy in the detectors on earth, a constraint on the mass can be directly
obtained.  Such a burst was observed in the IMB \cite{ref:IMB-SN87A} and
Kamiokande~II \cite{ref:Kamio-SN87A} detectors from SN1987A in the Large
Magellanic Cloud at a distance of $52$~kpc.  Useful limits are complicated to
obtain because of uncertainties in neutrino production models and because of
the low statistics ($8$ events from IMB and $11$ from Kamiokande~II), but
upper bounds of $M_\nu\: \lsim\: 30$~eV from SN1987A were obtained
from detailed analyses of the observations and astrophysical
constraints \cite{ref:SN87A-mass-kolb-turner,ref:SN87A-mass-derujula}.  With
the improvements in both tonnage and energy thresholds of contemporary
neutrino observatories, another happy occurrence like SN1987A may provide
additional direct $\nuebar$ mass constraints.


\subsection{Double Beta Decay}

\newcommand{\betabeta}{0\nu\beta\beta}
\newcommand{\betabetanunu}{2\nu\beta\beta}
\newcommand{\Mzero}{\mbox{$\cal M$}_{0\nu}}
\newcommand{\Mtwo}{\mbox{$\cal M$}_{2\nu}}

\newcommand{\mnutwo}{\mbox{$\left<M_\nu\right>^2$}}
\newcommand{\mnu}{\mbox{$\left<M_{\nu}\right>$}}

Unique amoung neutrino experiments, neutrinoless double beta decay
($\betabeta$) occurs only if the neutrino is not only massive but a Majorana
particle \cite{ref:pandr}.  $\betabeta$, the transition $(N,Z) \ra (N-2,Z+2) + e^-
+e^-$, occurs in any case when the mass of the parent nucleus $(N,Z)$ is greater than
the mass of the daughter nucleus $(N-2,Z+2)$, but the experimentally most
interesting cases are when the intermediate state $(N-1,Z+1)$ has a mass 
greater than that of the parent nucleus, supressing the single beta decay
$(N,Z) \ra (N-1,Z+1) + e + \nu$.  Sensitivity to the
Majorana mass arises because the righthanded antineutrino emitted in the
first $n \ra p + e^- + \nuebar$ transition must be reabsorbed as a
lefthanded neutrino in $\nue + n \ra p + e^-$. This reabsorption requires that
$\nuebar = \nue$, and for helicity reasons is suppressed
by a factor $M_{\nu}/E_{\nu}$, where $E_{\nu}$ is the energy of the
neutrino in the intermediate state.  As a result, the $\betabeta$ lifetime is
\begin{equation}
\left( T_{1/2, \betabeta} \right)^{-1} = G_{0\nu} | \Mzero |^2 \mnutwo \; ,
\end{equation}
where
\begin{equation}
\mnu \; = \sum_i U_{ei}^2 M_i \; .
\end{equation}
The neutrino mass eigenstates, whose masses are $M_i$, must now be Majorana particles.
Righthanded charged currents may affect the decay rate, but nonvanishing neutrino mass
is required for  $\betabeta$ to take place \cite{b30a}. 
$G_{0\nu}$ is a phase space factor and $\Mzero$ is the nuclear
matrix element.  Two neutrino double beta decay ($\betabetanunu$) may also
take place as an allowed process, and 
\begin{equation}
\left( T_{1/2, \betabetanunu} \right)^{-1} = G_{2\nu} | \Mtwo |^2
\end{equation}
gives the rate.

All experiments searching for $\betabeta$ rely on the measurement of the
total energy of the two final state electrons, which must equal the Q value of the
decay: $T_{e_1^-}+T_{e_2^-} = Q$.  Experiments fall into two broad catagories:
calorimeteric searches and tracking searches\footnote{Radio chemical experiments, in
which an excess of $\beta\beta$
daughters in old elemental deposits is used to measure the total
$\betabeta + \betabetanunu$ rate, are an exception \cite{ref:radio}.}.  
Calorimetric experiments use very high energy resolution to search for a peak in
the $T_{e_1^-}+T_{e_2^-}$ spectrum against a constant background
caused primarily by the Compton scattering of energetic gamma rays
from radioisotope contaminants in the experimental apparatus.  For a
detector with $N \times 10^{23} \; \beta\beta$ decay candidates, energy
resolution $\Delta E$ (in keV), and $B$ background counts per keV in
measuring time $t$ years, the halflife sensitivity is \cite{ref:bandv}
\begin{equation}
T_{1/2} \sim ln 2 \times 10^{23} \; y \left( \frac{Nt}{B \Delta E}
 \right)^{1/2}.
\end{equation}
The most sensitive experiments use enriched $^{76}$Ge cryogenic detectors
and attain total energy resolutions of 1.5 keV around $Q=2041$ MeV.
Germanium \cite{ref:gegenius,ref:igex,ref:ge}, CdTe \cite{ref:cdte}
and Te \cite{ref:milanote130} detectors have the 
source as the detection medium,
giving an efficiency approaching 100\% for a system with a large
number of emitters ($10^{26}$ in the case of the largest germanium experiment).
Detectors of 10 kg reach background limits of a few counts per year,
probing the lifetime range above $10^{25}$ yr.

Tracking type detectors identify both electrons via their ionization
deposition in a tracking medium and use either the total energy
deposition or the bending in a magnetic field (or both) to determine
the energy of each electron.  Identification of the two electrons
provides a powerful disciminant against backgrounds from $\gamma$,
$\alpha$ and $\beta$ emitters in the apparatus.  The use of $^{136}$Xe 
\cite{ref:xecit,ref:milano}
as a fill gas in a tracking type system gives high efficiency for a
large number of emitters ($4 \times 10^{25}$ in the case of the
largest experiment) by using the source as the detection medium. Other experiments 
\cite{ref:beijing,ref:arnold,ref:ca48,ref:mo100nd150,ref:nemomo100,ref:morend150,ref:osakamo100}
suspend the $\betabeta$ emitter in the tracking volume.  This limits the size of the
source, as the source must be thin enough for the electrons to escape without
significant energy loss in the source material.

The most sensitive experiments currently probe the lifetime range of
$10^{23}$ to $10^{25}$ years, as summarized in Table~\ref{ta:betabeta} for the most
experimentally attractive isotopes.  None yield an
observation of $\betabeta$ and extraction of an upper limit for $\mnu$ is
hampered by the complexity in calculating $\Mzero$.  Two general methods
are used in the computation of this nuclear matrix element 
\cite{ref:matrix}: the shell
model and the quasiparticle random phase approximation (QRPA). Both
give comparable results at the factor of five level.  The QRPA is
sensitive to the particle-particle coupling, which is poorly known and
must be extracted from positron decay experiments.  In the allowed
range, the QRPA calculations indicate $\betabeta$ may be completely
suppressed by the nuclear matrix elements in some cases. Use of both
methods gives rough agreement for the observed $\betabetanunu$ rates.
Overall, the limit on $\mnu$ from $\betabeta$ lies around $\mnu \, <1$eV.

\begin{table}
\begin{center}
\begin{tabular}{|l l r c|}
\hline
Isotope    & Experiment & $T_{1/2,\betabeta}$ (yr) & $\mnu_{u.l.}$(eV)
\\
\hline
$^{48}$Ca & HEP Beijing \cite{ref:beijing} & $>1.1 \times 10^{22}$, 68\% c.l. & 23-50 \\ 
$^{76}$Ge & Heidelberg/Moscow \cite{ref:gegenius} & $>5.7 \times 10^{25}$, 90\% c.l. &
0.2-0.8 \\ 
          & IGEX \cite{ref:igex}   & $>0.8 \times 10^{25}$, 90\% c.l. & \\ 
$^{82}$Se & Irvine  & $>2.7 \times 10^{22}$, 68\% c.l. &
4-14 \\ 
          & NEMO 2 \cite{ref:arnold}  & $>9.5 \times 10^{21}$, 90\% c.l. & \\ 
$^{96}$Zr & NEMO 2 \cite{ref:nemo2zr}  & $>1.3 \times 10^{21}$, 90\% c.l. & -
\\ 
$^{100}$Mo & LBL \cite{ref:igex}    & $>2.2 \times 10^{22}$, 68\% c.l. & 3-111\\ 
           & UCI \cite{ref:ucimo100}  & $>2.6 \times 10^{21}$, 90\% c.l. & \\ 
           & Osaka \cite{ref:osakamo100}   & $>2.8 \times 10^{22}$, 90\% c.l. & \\ 
           & NEMO 2 \cite{ref:nemomo100}           & $>5 \times 10^{21}$, 90\% c.l. & \\ 
$^{130}$Te & Milano \cite{ref:milanote130}          & $>7.7 \times 10^{22}$, 90\% c.l. & 2-5\\ 
$^{136}$Xe & Caltech/PSI/Neuchatel \cite{ref:xecit}& $>4.4 \times 10^{23}$, 90\% c.l. & 2-5\\ 
$^{150}$Nd & UCI \cite{ref:morend150}      & $>1.2 \times 10^{21}$, 90\% c.l. & 5-6 \\
\hline
\end{tabular}
\caption{\label{ta:betabeta} Current experimental limits on
    $\betabeta$ decay and upper limits on $\mnu$.}
\end{center}
\end{table}

The present generation of operating experiments will achieve
sensitivities of $T_{1/2, \betabeta} \sim 10^{26}$ yr, or $\mnu \:\lsim \: 0.3$ eV. 
Ambitious plans are underway to push this limit even further:
the GENIUS \cite{ref:genius} experiment seeks to probe $\mnu \,< 0.1$ eV through the
use of several tons of enriched $^{76}$Ge detectors. The CUORE
experiment will use a cryogenic TeO$_2$ detector to achieve very high
energy resolutions for large masses \cite{ref:curore}. If, for example, the neutrinos
have a quasi-degenerate mass spectrum (see Section~\ref{sect:scena}), the coming
experiments may cast light on the Majorana nature of the neutrino.


\section{NEUTRINO MASS SCENARIOS}\label{sect:scena}

We have discussed the fairly strong evidence for neutrino mass from the behavior of
the atmospheric neutrinos, the major further hint of mass from the behavior of the
solar neutrinos, and the unconfirmed additional hint from the LSND experiment. To what
neutrino mass scenarios do these hints point?

For simplicity, one would like to assume that there are just three neutrinos of
definite flavor, $\nu_e$, $\nu_\mu$, and $\nu_\tau$, and just three corresponding
neutrinos of definite mass, $\nu_1$, $\nu_2$, and $\nu_3$. However, for simplicity,
one would also like to assume that the solar, atmospheric, and LSND oscillations each
involve just a single oscillation ``frequency'', $\delta M^2$. But when one makes the
latter assumption, the inferred frequencies of the three oscillations cannot all be
accommodated in terms of the masses of just three neutrino mass eigenstates. For,
these frequencies are, as we have seen, approximately as summarized by Table \ref{t1}.
Now, if there are only three mass eigenstates $\nu_i$, then there are only three
different mass splittings $\delta M^2_{ij} \equiv M^2_i - M^2_j$, and these three
splittings obviously satisfy
\beq
\sum_{\mbox{\scriptsize Splittings}} \delta M^2_{ij} = (M^2_3 - M^2_2) + (M^2_2 -
M^2_1) + 
(M^2_1 - M^2_3) = 0 \;.
\label{eqs1}
\eeq
\begin{table}     
\begin{center}
\vspace{.2in}
\begin{tabular}{|c|c|} 
\hline 
\raisebox{0pt}[12pt][6pt]{Oscillating Neutrinos} & 
  \raisebox{0pt}[12pt][6pt]{Required $\left| \delta M^2 \right|$\,(eV$^2$)} \\
\hline
\raisebox{0pt}[12pt]{Solar} & 
  \raisebox{0pt}[12pt]{$10^{-10}$ or $10^{-5}$} \\
Atmospheric & $10^{-3}$ to $10^{-2}$ \\
LSND & $10^{-1}$ to $10^{+1}$ \\ 
\hline
\end{tabular}   
\caption{Mass splittings required by the three hints of oscillation.}
\label{t1}
\end{center}
\vspace*{3pt}
\end{table}
But, as we note from Table~\ref{t1}, the splittings required by the three oscillations
are, respectively, of three different orders of magnitude. Thus, they cannot possibly
add up to zero, as demanded by \Eq{s1}, no matter what sign we assign to each of them.
Hence, the three oscillations cannot all be explained in terms of just three
neutrinos. To accommodate all three of these oscillations under the assumption that
each oscillation is described by a single $\delta M^2$, we must introduce (at least) a fourth neutrino.
Since we know from the width of the Z boson that only three species of neutrinos have
normal weak interactions, this extra, fourth neutrino must be a sterile neutrino
$\nu_S$ \cite{b10}.

Since no direct evidence for any sterile neutrino has been seen, one wonders whether
the simple assumption that each oscillation involves but a single $\delta M^2$ is
false. Suppose, for example, that there are only three neutrinos, with masses such
that $M^2_3 - M^2_2 \equiv \delta M^2_{\mbox{\scriptsize Big}} \gg M^2_2 - M^2_1
\equiv \delta M^2_{\mbox{\scriptsize Small}}$. Could it be that the oscillation of the
LSND neutrinos involves $\delta M^2_{\mbox{\scriptsize Big}}$, the flavor conversion
of the solar neutrinos involves $\delta M^2_{\mbox{\scriptsize Small}}$, and the
oscillation of the atmospheric neutrinos involves a mixture of both $\delta
M^2_{\mbox{\scriptsize Big}}$ and $\delta M^2_{\mbox{\scriptsize Small}}$? 
Then an analysis of the atmospheric data assuming (erroneously) that only one $\delta
M^2$ is involved might find a value intermediate between those corresponding to the
LSND and solar effects, as observed. Nevertheless, in this scenario, there are only
three underlying neutrinos.

A number of analyses have attempted to exploit this possibility \cite{b31}. These
attempts are clearly important, and should continue. However, the present authors have
not seen any attempt to describe all three oscillations with just three neutrinos
which is not at least somewhat inconsistent with some of the data. This being the
case, we assume here that if one is to describe all three of the oscillations, at
least four neutrinos are needed. 

\begin{figure}    
\center
\epsfxsize=1.0\textwidth\epsfbox{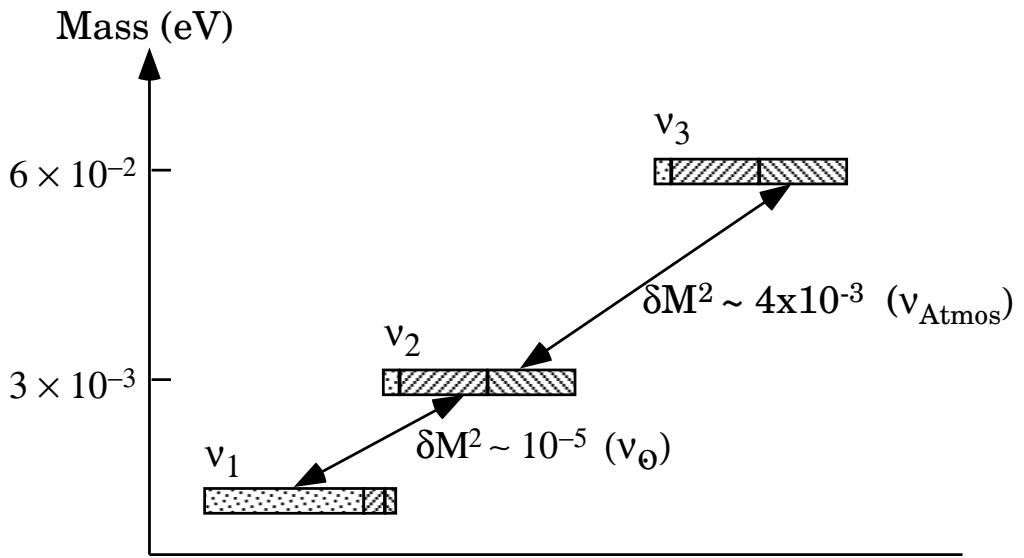}
\caption{A three-neutrino mass hierarchy that accounts for the oscillations of
atmospheric and solar neutrinos. The neutrinos $\nu_1$, $\nu_2$, and $\nu_3$ are  mass
eigenstates. The rough flavor content of each of these is indicated by dotting and
hatching: The $\nu_e$ fraction of a mass eigenstate is dotted, the $\nu_\mu$ fraction
is indicated by right-leaning hatching, and the $\nu_\tau$ fraction by left-leaning
hatching \cite{b32}. }
\label{f7}
\end{figure}

If one is reluctant to introduce a (light) sterile neutrino, then he can proceed by
provisionally setting aside the so-far unconfirmed LSND oscillation, and attempting to
explain only the atmospheric and solar neutrino data in terms of oscillation. One can
then have the three-neutrino mass spectrum depicted in Figure~\ref{f7}, which assumes
the behavior of the solar neutrinos is due to the small-angle MSW effect. 
The masses of the neutrinos in Figure~\ref{f7} form a hierarchy, with $M_3 \gg M_2 \gg
M_1$. The splitting $M^2_3 - M^2_2 \simeq M^2_3 \simeq 4 \times 10^{-3} \mbox{eV}^2$
gives the atmospheric neutrino oscillation, while the splitting $M^2_2 - M^2_1 \simeq
M^2_2 \simeq 10^{-5} \,\mbox{eV}^2$ yields the MSW effect in the sun.

The hierarchy in Figure~\ref{f7} is the lightest one which gives the $\delta M^2$
values called for by the atmospheric and solar oscillations. Since oscillations
determine only mass splittings, and not actual masses, one, of course, can increase
the masses in Figure~\ref{f7} without affecting the oscillations so long as one keeps
the $\delta M^2$ values fixed. For example, if neutrinos contribute significantly to
$\Omega_M$, then perhaps we have a quasi-degenerate neutrino mass spectrum with $M_1
\simeq M_2 \simeq M_3 \simeq (1-2)\,\mbox{ eV}$ and the splittings $\delta M^2$ as in
Figure~\ref{f7}.

Suppose the behavior of the solar neutrinos is due to oscillation in vacuum, rather
than to the MSW effect. Then the masses of the neutrinos in Figure~\ref{f7} must be
changed to give the small $\delta M^2 \,(\sim 10^{-10} \mbox{eV}^2)$ that corresponds
to solar neutrino vacuum oscillations. Also, their flavor content must be changed to
reflect large solar neutrino mixing $(\sin 2 \theta \sim 0.8)$. However, a neutrino
mass-and-mixing scenario that corresponds to the solar and atmospheric observations is
easily obtained \cite{b33}.

\begin{figure}    
\epsfxsize=1.0\textwidth\epsfbox{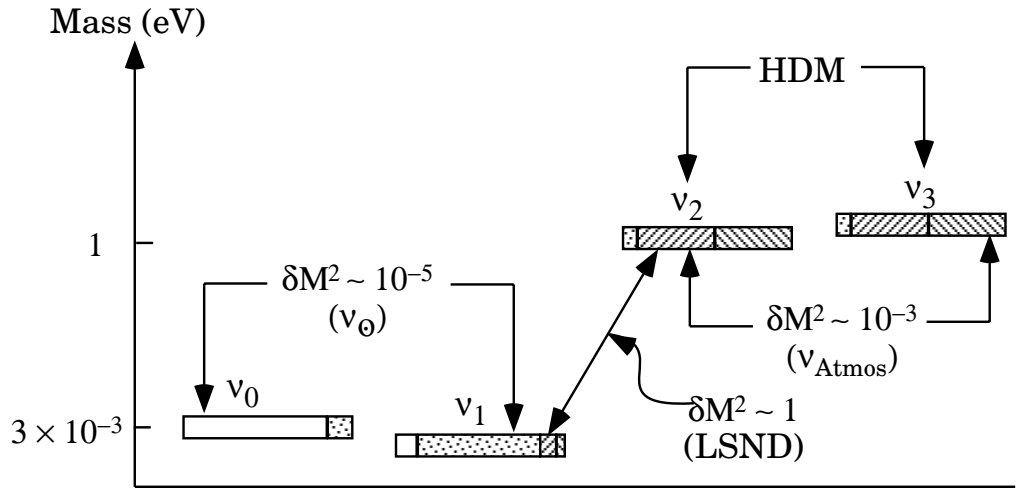}
\caption{A four-neutrino scenario that accounts for the oscillations of the
atmospheric, solar, and LSND neutrinos. The scenario contains the mass eigenstates
$\nu_0 - \nu_3$. The $\nu_e$, $\nu_\mu$, and $\nu_\tau$ fractions of these mass
eigenstates are indicated as in Figure~\ref{f7}, and the $\nu_S$ fraction is shown
as a clear region. The mass eigenstate $\nu_0$ is largely a sterile neutrino $\nu_S$.
The $\sim 1$ eV neutrinos in the scheme make a significant contribution to Hot Dark
Matter.}
\label{f8}
\end{figure}

If one is willing to include a $\nu_S$ among the light neutrinos, then all three hints
of oscillation can be accommodated. One neutrino mass scenario which accommodates them
is shown in Figure~\ref{f8} \cite{b34}. In this scenario, the mass spectrum includes
two $\sim 1$ eV mass eigenstates, $\nu_2$ and $\nu_3$, split by a $\delta M^2 \sim
10^{-3} \mbox{ eV}^2$ that produces the atmospheric neutrino oscillation. 
It also includes two much lighter mass eigenstates, $\nu_0$ and $\nu_1$, split by a
$\delta M^2 \sim 10^{-5} \mbox{ eV}^2$ that yields the MSW effect in the sun. However,
as we see from Figure~\ref{f8}, in this scheme the MSW effect converts a $\nu_e$
into the largely sterile neutrino $\nu_0$ rather than into a $\nu_\mu$ or $\nu_\tau$.
Thus, there is no $\nu_\mu$ or $\nu_\tau$ component in the solar neutrino flux.
Finally, the splitting $\delta M^2 \sim 1 \mbox{ eV}^2$ between the heavy pair $\nu_2$
and $\nu_3$, and the light pair $\nu_0$ and $\nu_1$, produces the LSND oscillation.
The heavy pair in this scenario has been given the smallest average mass, $\sim 1$eV,
which can still yield the $\delta M^2 \sim 1 \mbox{ eV}^2$ called for by LSND. 
Interestingly, with this $\sim 1$ eV mass, the members of the pair make a significant
Hot Dark Matter contribution to the mass density of the universe, but without
violating the cosmological upper bound of \Eq{HDM} on neutrino mass. 

In a theory containing the see-saw mechanism for neutrino mass, a sterile neutrino
occurs naturally, as we have seen in Section~\ref{sec2}. However, this sterile
neutrino is very heavy, with a mass in the multi-GeV range or higher. In general,
theories of neutrino mass do not predict {\em light} sterile neutrinos, such as the
one in Figure~\ref{f8}. However, even if one does not include LSND, 
the large mixing angles observed in atmospheric, and perhaps solar, neutrino
oscillation also suggest that light sterile neutrinos may exist. The small mixing
angles observed in the quark sector of the Standard Model lead us to expect naively
that the mixing angles between active species of neutrinos are also small. However,
there would be no such expectation for mixing between active and sterile neutrinos.
Indeed, appealing theories have been constructed in which, as a result of a symmetry,
the mixing between an active neutrino and a sterile one is automatically maximal
\cite{b35}.

Firm establishment of the existence of a light sterile neutrino would be a 
groundbreaking result. Thus, it is very important to confirm or disprove each of the
three present hints of oscillation, and to thoroughly examine the question of whether
or not these three oscillations, if all confirmed, together require the existence of a
fourth, sterile neutrino. It is also important to carry out experiments which can tell
whether solar or atmospheric neutrino oscillation, or both, involve a sterile
neutrino.

The neutrino-mass scenarios we have discussed are just examples. Other scenarios are
possible \cite{b33}. As discussed in our concluding section, the various neutrino mass
and mixing possibilities will be tested through future experiments.


\section{CONCLUSIONS}

To conclude this breathless sprint through the rich world of neutrino
oscillation and neutrino mass physics, we consider the possible paths through
the maze of revealing measurements that are expected in the next five or ten
years and where this information might point experimenters and theorists in
the long term.  Some of the mainstream experimental questions to be answered
are summarized in Table~\ref{ta:rouse}.  {\em Caveat emptor}: if the rapid
changes in our knowledge in the past ten years is any indication, the phase
space for surprises outside of the scope of these questions may be large.  We
consider, in turn, probes of our three neutrino oscillation hints: solar
neutrinos, atmospheric neutrinos and the LSND signal.

\begin{table}
\begin{tabular}{p{0.25\textwidth}|p{0.3\textwidth}|p{0.2\textwidth}|p{0.12\textwidth}}
{\raggedright Future observation}   
& {\raggedright Solar $\nu$ oscillation} 
& {\raggedright Atmospheric $\nu$ oscillation} 
& {\raggedright LSND $\nu$ oscillation} \\ \hline\hline
{\raggedright $\sim 1$~eV $\nu_e$ in tritium decay  }
& {\raggedright If solar $\nu$ osc., then at least
2-fold degeneracy with $\delta M^2 <$ 10$^{-5}$  }
& {\raggedright -} & 
{\raggedright $\dmsq$ gives us masses of two eigenstates} \\ 
\hline
$0\nu\beta\beta$ at $\sim 0.1$eV ($\Rightarrow~\nu = \overline{\nu}$) 
&  -   & May be relevant & $\nu_e \to \overline{\nu_e}$ contributes?\\
\hline
SNO NC/CC as expected   & Solar $\nu_e \to \nu_\mu~{\rm or}~\nu_\tau$ &
NC deficit implies only one $\nu_s$ which participates here   &- \\ 
\hline
SNO NC/CC deficit & $dN/dE_{\nu}$ distortion proves oscillations, 
and therefore $\nu_s$  
& NC deficit here implies another $\nu_s$ & - \\
\hline
Deep BOREXINO $^7$Be deficit & 
 Big $dN/dE_{\nu}$ distortion confirms small angle MSW   &  -       & -  \\
\hline
BOREXINO seasonal variation & $\nu_e$ vacuum oscillations &  - & -\\
\hline
$\nuebar$ disappearance in KAMLAND & 
Large angle MSW (day/night or small $dN/dE_{\nu}$ distortion would confirm) 
&  -    &  -  \\ 
\hline
Results of long baseline experiments (MINOS, K2K, CERN) & - &
Confirm $\nu_\mu \to \nu_\tau$ if see $\tau$ appearance  &  -  \\
\hline
Confirmation and measurement of LSND $\dmsq$ by BooNe, I-216 & 
\multicolumn{3}{p{0.69\textwidth}}{\centering Three signatures suggest three distinct
  mass splittings, and thus a fourth light (sterile) neutrino needed} \\
\end{tabular}
\caption{\label{ta:rouse} Some possible outcomes of future observations
  are listed.  Next to each possible outcome are listed consequences 
  or provocative additional experimental signatures related to the solar,
  atmospheric or LSND neutrino oscillations.}
\end{table}

The low $\delta M^2$ expected to be responsible for the flavor conversion of
solar neutrinos precludes accelerator-based probes of this signal in the near
future.  However, on the observational front, both SNO and BOREXINO will
begin data taking.  BOREXINO will measure the $^7$Be flux and search for
seasonal variations in the solar neutrino flux.  Observation of a deep
deficit of $^7$Be flux would indicate that solar neutrinos oscillate
according to the small angle MSW solution or in vacuum.  This would be
confirmed by the observation at SNO or SuperKamiokande of a deviation from
the neutrino energy distribution in the absence of oscillations.  Observation
of a seasonal variation at BOREXINO would mean the oscillation depends on the
earth-sun distance, indicating vacuum oscillations play a role in solar
neutrino behavior.  Perhaps the most exciting probe is SNO's measurement of
the neutral current solar neutrino interactions.  Muon and tau neutrinos have
neutral current interactions of the same strength as electron neutrinos.
Thus, if the missing solar electron neutrinos oscillate into another active
neutrino flavor, either muon or tau, the rate for neutral current
interactions of solar neutrinos will be the same as in the absence of
oscillation.  However, if the missing electron neutrinos are oscillating into
sterile neutrinos, the neutral current rate will be reduced.  If such a
reduced rate is observed and if additional confirmation of the oscillation
hypothesis for solar neutrino disappearance is obtained, such as a large
day/night effect or anomalous solar neutrino energy spectrum, the case for
$\nu_e \to \nu_s$ oscillations will become increasingly compelling.  If the
large angle MSW solution is indicated by small distortions of the solar
neutrino energy spectrum and a large day/night effect, then the KAMLAND
reactor experiment will take center stage with the first possibility for a
terrestrial confirmation of the solar neutrino oscillation hypothesis.

The path to verifying the oscillation hypothesis for the atmospheric neutrino
anomaly seems to clearly point to the next generation of long-baseline
accelerator experiments, MINOS, K2K and CERN to Gran Sasso.  Unless $\delta
M^2$ is at the lowest end of the range suggested by the SuperKamiokande data,
all three programs should have access to observing the $\nu_\mu$
oscillations; however, since $L/E$ in all these experiments gives a peak
sensitivity at $\delta M^2\sim10^{-2}$, there is a significant possibility
that these experiments will not be able to measure $\delta M^2$ by observing
a complete oscillation cycle.  Therefore, a low $\delta M^2$ may require
either substantial beam redesign or a different baseline, which may in turn
drive a new generation of long-baseline experiments.  Since the reactor
oscillation experiments (and the SuperKamiokande data itself) rule out the
$\nu_\mu \to \nu_e$ interpretation of the atmospheric data, a premium is
placed on the ability to detect $\tau$'s or neutral current $\nu_\tau $
interactions in order to determine whether the atmospheric neutrinos
oscillate to $\nu_\tau$ or to a sterile neutrino.  Here, the higher energy
experiments, MINOS and the CERN-Gran Sasso beam, will have an advantage over
K2K, which cannot produce $\tau$ leptons.  
Should CP violation in neutrino oscillations be observed in these long
baseline experiments, this would imply that at least three neutrino mass
eigenstates have splittings, $\delta M^2$, large enough to be within the
sensitivity of these experiments.  This could be interpreted within the three
neutrino framework as favoring an explanation for the solar neutrino deficit
other than oscillation with a very small $delta M^2$, or it could imply the
presence of a fourth, sterile neutrino.

New experiments, BooNE at Fermilab and I-216 at CERN, should be able to
conclusively confirm or refute the LSND signal, and measure $\delta M^2$ and
the mixing if the effect is observed.  If the effect persists, it is possible
the electron neutrino lies within the reach of a kinematic measurement.
Kinematic techniques must be employed in order to resolve the overall scale
of neutrino masses since oscillation is only sensitive to splittings.  At this
time, the best one can hope for is a mass sensitivity of $0.1$~eV for $\nue$,
which would begin to probe mass splittings favored by the LSND result.  If
the electron neutrino were shown to have a mass $\sim0.1$~eV, the solar
neutrino oscillation would indicate there are two nearly mass degenerate
neutrinos with $\delta M^2 < 10^{-5}$ eV$^2$.  A very interesting and
important experiment would be one which could resolve $\nu_\mu$ or $\nu_\tau$
masses at the $\delta M^2 \sim 1$ eV$^2$ level.

The Majorana or Dirac nature of the neutrino remains an open question.  We
can hope for a mass sensitivity of $0.1$~eV for Majorana electron neutrinos
in the coming years which, if observed, indicates the LSND oscillation could
be at least partially attribtued to $\nu_e \to \overline{\nu_e}$.  Direct
probes of the question of the Majorana nature of the other neutrinos 
is likely to be unresolved for the forseeable future.

Implicit in many of these outcomes is very exciting physics beyond the
Standard Model.  The survival of all three current hints, discovery of CP
violation inconsistent with only three neutrinos, or direct probes
for solar or atmospheric neutrino ``disappearance'' may indicate the presence
of a new particle, a light sterile neutrino.
Understanding its relationship to our current picture of exactly three
generations of quarks, leptons and light neutrinos will prove a major
challenge, both theoretically and experimentally, as the generational
structure of fermions is probed at the energy frontier of accelerator
experiments. 


In summary, current experimental results strongly indicate
neutrinos have mass and the different lepton flavors
mix.  While lepton family number appears to be violated, the question
of total lepton number conservation remains open.  Currently-operating
and soon-to-start experiments will take important steps toward
completing the neutrino picture, but some large issues most likely will
not soon be resolved.  We expect that neutrino physics will remain a central
focus of both experimental and theoretical endeavor for a long time to come.

\subsection*{Acknowledgements}

We would like to thank John Bahcall, Robert Bernstein, Felix Boehm, David
Caldwell, Martin Deutsch, George Fuller, Tom Gaisser, Steve Geer, Maury
Goodman, Giorgio Gratta, Donald Groom, John Learned, Eugene Loh, John
LoSecco, Bill Louis, Geoff Mills, Stephen Parke, Serguey Petcov, Jon Rosner,
Henry Sobel, Jack Steinberger, Petr Vogel and Lincoln Wolfenstein for helpful
and stimulating discussions during the formulation of this review.  We owe a
special debt to Susan Kayser for her heroic efforts in helping to prepare
this document.  One of us (BK) would like to acknowledge the hospitality of
the Aspen Center for Physics where there were many opportunities to exchange
ideas about neutrino mass.  This work was supported by the US Department of
Energy and the Alfred P. Sloan Foundation.







\end{document}